\documentclass[trackchanges,twocolumn,table]{aastex63}
\usepackage[utf8]{inputenc}
\usepackage{graphicx}
\usepackage{hyperref}
\usepackage{amssymb}
\usepackage{amsmath}
\usepackage{amsfonts}
\usepackage{calc}

\shorttitle{Random Walk Model for Halo Concentrations}
\shortauthors{Johnson, Benson, \& Grin}

\submitjournal{ApJ}

\begin{document}
\title{A Random Walk Model for Dark Matter Halo Concentrations}

\correspondingauthor{Andrew Benson}
\email{abenson@carnegiescience.edu}

\author{Turner Johnson}
\affiliation{Department  of  Physics  and  Astronomy,  Haverford  College, 370  Lancaster  Avenue,  Haverford,  Pennsylvania  19041,  USA}

\author{Andrew J. Benson}
\affiliation{Carnegie Observatories, 813 Santa Barbara Street, Pasadena, CA 91101, USA}

\author{Daniel Grin}
\affiliation{Department  of  Physics  and  Astronomy,  Haverford  College, 370  Lancaster  Avenue,  Haverford,  Pennsylvania  19041,  USA}

\begin{abstract}
 For idealized (spherical, smooth) dark matter halos described by single-parameter density profiles (such as the NFW profile) there exists a one-to-one mapping between the energy of the halo and the scale radius of its density profile. The energy therefore uniquely determines the concentration parameter of such halos. We exploit this fact to predict the concentrations of dark matter halos via a random walk in halo energy space. Given a full merger tree for a halo, the total internal energy of each halo in that tree is determined by summing the internal and orbital energies of progenitor halos.  We show that, when calibrated, this model can accurately reproduce the mean of the concentration--mass relation measured in N-body simulations, and reproduces more of the scatter in that relation than previous models. We further test this model by examining both the autocorrelation of scale radii across time, and the correlations between halo concentration and spin, and comparing to results measured from cosmological N-body simulations. In both cases we find that our model closely matches the N-body results. Our model is implemented within the open source {\sc Galacticus} toolkit.
\end{abstract}


\section{Introduction}

It has long been established that dark matter halos forming in a universe dominated by cold dark matter have a ``universal'' density profile \citep{2008ApJ...685..739B}. This is often described by the NFW functional form \citep{navarro_universal_1997}, or, more recently, by the Einasto profile (\citeauthor{einasto_construction_1965}~\citeyear{einasto_construction_1965}; see, for example, \citealt{navarro_inner_2004}). The universality has been explained using a radial orbit instability, as in \cite{navarro_universal_1997}, or adiabatic contraction near the peaks of Gaussian density fields, as in \cite{2010arXiv1010.2539D}.

Given a halo mass, $M_\mathrm{v}$, defined as the mass within some sphere or isodensity surface enclosing a given density contrast, the NFW profile is characterized by a single parameter, the scale radius, $r_\mathrm{s}$. This variable is often parameterized in terms of the concentration, $c=r_\mathrm{v}/r_\mathrm{s}$, where $r_\mathrm{v}=(3 M_\mathrm{v}/4 \pi \Delta_\mathrm{v} \bar{\rho})^{1/3}$ is the virial radius of the halo, encompassing the halo mass. Here $\bar{\rho}$ is the mean density of the universe, and $\Delta_\mathrm{v}$ is a suitably-chosen virial density contrast.

Understanding how the concentration is related to halo properties (such as the halo mass, formation time, etc.) is important for a number of theoretical and observational reasons. For example:
\begin{itemize}
    \item more concentrated halos are expected to be able to survive longer in tidal fields \citep{jiang_satgen_2020}, thereby affecting the number of surviving subhalos found around galaxies and within clusters;
    \item concentration has been shown to be a good predictor of galaxy sizes in hydrodynamical simulations of galaxy formation \citep{jiang_is_2019};
    \item concentration directly affects a halo's cross section for gravitational lensing, making it an important ingredient in analyses of strong lensing systems which aim to constrain the particle nature of dark matter \citep{gilman_warm_2020,gilman_constraints_2020}.
\end{itemize} 

As first shown by \citep{navarro_structure_1996}, this scale radius is closely correlated with the formation history of a halo, with earlier-forming halos being more concentrated (larger $c$, smaller $r_\mathrm{s}$) than halos forming later at fixed halo mass. This correlation of concentration with the assembly history of a halo was further explored by \cite{2001MNRAS.321..559B}, \cite{wechsler_concentrations_2002} and \cite{zhao_accurate_2009}.

More recently, \citeauthor{ludlow_mass-concentration-redshift_2016}~(\citeyear{ludlow_mass-concentration-redshift_2016}; see also \citealt{ludlow_mass-concentration-redshift_2014}) proposed a model in which a halo's scale radius is determined by the epoch at which a certain fraction of the final halo mass was first assembled into progenitor halos above a certain mass threshold. This approach improves upon those described above by accounting for the effects of the specific merger history of each halo on its concentration, rather than considering just the typical assembly history for halos of a given mass. As such, the approach of \cite{ludlow_mass-concentration-redshift_2016} is able to capture halo-to-halo variations in concentration at fixed halo mass, thereby explaining the origin of the majority of the scatter in the concentration--mass relation.

\cite{ludlow_mass-concentration-redshift_2016} demonstrated that this model could accurately predict the concentrations of individual N-body halos given their formation histories (i.e. the set of progenitor halos). This clearly demonstrated that the concentration of a given halo is closely connected to the corresponding distribution of its progenitor halos.

\cite{benson_halo_2019} applied the \cite{ludlow_mass-concentration-redshift_2016} semi-analytic model for concentrations to merger trees constructed using the algorithm proposed by \cite{parkinson_generating_2008}, and used this to predict the distribution of concentrations for $z=0$ halos in different mass intervals. \cite{benson_halo_2019} found that, while the mean of the predicted distribution of concentrations accurately matched that measured from N-body simulations (specifically the COCO simulations of \citealt{hellwing_copernicus_2016}), the scatter in the distribution was significantly smaller than that of the N-body distribution of concentrations, even after introducing some dependence of halo formation histories on large scale environment.

In this work, we hypothesize that at least some of this ``missing'' scatter in the distribution of concentrations must be due to the fact that the \cite{ludlow_mass-concentration-redshift_2016} model does not make use of the entirety of the information available in a merger tree, but instead makes use of what is essentially a summary statistic (i.e. the epoch at which a certain mass fraction is assembled into progenitors above a certain mass). \cite{wang_concentrations_2020} have explored the effects of major and minor mergers on halo concentrations measured from N-body simulations, finding that ``merger events induce lasting and substantial changes in halo structures.'' \cite{wang_concentrations_2020} conclude that minor mergers are a source of irreducible scatter in the concentration--mass relation, bolstering our hypothesis.

In this work we therefore develop a model for predicting scale radii which takes into account the entire structure of a merger tree and which we therefore expect to capture more of the scatter in concentration at fixed halo mass.

The inspiration for this model is the work of \cite{vitvitska_origin_2002}, who construct a random walk model for the spin parameters (i.e. the dimensionless measure of the internal angular momenta) of halos. Briefly, they assume that whenever two halos merge, the spin angular momentum of the merged system is equal to the sum of the spin angular momenta of the two merging halos, plus the orbital angular momenta of those halos at the point of merging (i.e. when the smaller halo first crosses the virial radius of the larger halo). This assumption is applied at each merging event in a tree to predict the evolution of spin along each branch. Recently, \cite{benson_random-walk_2020} updated this model and demonstrated that it can provide an excellent match to the distribution of spin parameters measured in N-body simulations, and also gives a reasonable reproduction of the correlation properties of spins across time.

In this work we apply this same approach to predicting halo scale radii and shapes by assuming that energy is similarly (approximately) conserved during halo mergers. This allows us to compute the internal energy of each halo in a merger tree, which we then map into scale radii for each halo. Models of this type take into account the entirety of a halo's merger history, and incorporate information from mergers across all mass ratios. Importantly, no artificial distinction is drawn between minor and major mergers---the mass ratio of mergers is treated as a continuum.

The remainder of this paper is organized as follows. In \S\ref{sec:model} we describe our model for halo scale radii, and how we constrain the parameters of this model. In \S\ref{sec:results} we present results for the distribution of concentrations in different mass intervals compared to measurements from N-body simulations, and explore predictions from our model for the correlation structure of scale radii across time, and correlations between scale radius and halo spin. Finally, in \S\ref{sec:conclusions} we give our conclusions. We include two appendices. In Appendix~\ref{app:unresolved} we describe how we model the effects of unresolved accretion, and demonstrate the validity of the approach through a resolution study. In Appendix~\ref{sec:posterior} we provide the full posterior distribution of our model parameters. The model developed in this work is implemented and available within the open source {\sc Galacticus} toolkit.

\section{Model for scale radii}\label{sec:model}

We begin by constructing merger trees using the algorithm of \cite{parkinson_generating_2008} with parameters given by \cite{benson_halo_2019}. As in \cite{benson_halo_2019}, we include a modification to the halo branching rates which captures the effects of large scale environment. All trees are rooted in a $z=0$ halo whose mass, $M_0$, is chosen from some distribution (we will describe the distribution of $z=0$ halo masses in each application of our model below), and are grown backward in time using the \cite{parkinson_generating_2008} algorithm until the halo mass along each branch reaches a pre-defined threshold, which we refer to as the mass resolution, $M_\mathrm{res}$. As in \cite{benson_random-walk_2020} we choose to set $M_\mathrm{res} = f_\mathrm{res} M_0$, with a fiducial fractional resolution of $f_\mathrm{res}=10^{-3}$. Any progenitor halos with mass less than $M_\mathrm{res}$ are therefore missing from our merger trees. The effects of these sub-resolution halos are not ignored however, as will be discussed below.

Our model is applicable to any combination of cosmological parameters. In this work we will calibrate our semi-analytic model to the COCO \citep{hellwing_copernicus_2016} N-body simulations which assume a cosmology of $(\Omega_\mathrm{m}, \Omega_\Lambda, \Omega_\mathrm{b},H_0/\hbox{km s}^{-1}\hbox{Mpc}^{-1},\sigma_8,n_\mathrm{s})=(0.272, 0.728,0.0446,70.4,0.81,0.967)$. In \S\ref{sec:autoCorrelation} we will test our model by comparing to the VSMDPL N-body simulation \citep{klypin_multidark_2016} which assumes a cosmology of $(\Omega_\mathrm{m}, \Omega_\Lambda, \Omega_\mathrm{b},H_0/\hbox{km s}^{-1}\hbox{Mpc}^{-1},\sigma_8,n_\mathrm{s})=(0.307,0.693,0.0482,67.77,0.823,0.96)$.

For specificity we will assume a \citeauthor{navarro_universal_1997}~(\citeyear{navarro_universal_1997}; ``NFW'') density profile throughout (although we explore the consequences of instead using an Einasto profile in Appendix~\ref{sec:profile}, where we show that the our qualitative conclusions are unaffected by the choice of density profile). The NFW profile is given by:
\begin{equation}
 \rho(r) = \frac{\rho_\mathrm{s}}{ (r/r_\mathrm{s})(1+r/r_\mathrm{s})^2}
\end{equation}
where $r_\mathrm{s}$ is the scale radius, and $\rho_\mathrm{s}$ is a normalization factor.

\subsection{The relation between energy and concentration}

Our model relies on determining the internal energy of a halo. We find the energy following the approach of \cite{cole_hierarchical_2000}, assuming the halo to be spherical, supported by isotropic velocity dispersion, and in Jeans equilibrium. The gravitational energy is given by:
\begin{equation}
 W = -{{\mathrm G}\over 2} \left[ \int_0^{r_\mathrm{v}} {M^2(r) \over r^2} \mathrm{d}r + {M^2(r_\mathrm{v}) \over r_\mathrm{v}}\right],
 \label{eq:energyPotential}
\end{equation}
where $M(r)$ is the mass enclosed within a sphere of radius $r$.

The kinetic energy is given by:
\begin{equation}
 T = 2 \pi \left[ r_\mathrm{v}^3 \rho(r_\mathrm{v}) \sigma^2(r_\mathrm{v}) + \int_0^{r_\mathrm{v}} \mathrm{G} M(r) \rho(r) r \mathrm{d}r \right],
 \label{eq:energyKinetic}
\end{equation}
where $\sigma(r)$ is the velocity dispersion at radius $r$ is determined from the Jeans equation:
\begin{equation}
 {\mathrm{d}(\rho\sigma^2) \over \mathrm{d}r} = - \rho(r) {\mathrm{G} M(r) \over r^2}.
 \label{eq:energySelfGravitational}
\end{equation}
\cite{cole_hierarchical_2000} derive the velocity dispersion at $r_\mathrm{v}$ by integrating the Jeans equation to $r=\infty$ assuming that the \cite{navarro_universal_1997} profile and Jeans equilibrium apply at all radii.

Importantly, by truncating the halo at $r_\mathrm{v}$ in this way the halo is not precisely in virial equilibrium: $2 T + W \ne 0$.

As was shown by \cite{cole_structure_1996}, the above estimates of the potential and kinetic energies of NFW halos result in virial ratios, $2T/|W|$, in good agreement with those found in N-body simulations. Specifically, when computed in this way, the virial ratio is independent of halo mass and depends only on the concentration parameter. Utilizing the VSMDPL N-body simulation Rockstar halo catalogs, which provide both the concentration and the virial ratio, we can compare these analytic expectations for halo energies to N-body results. The Rockstar algorithm is fully described by \cite{behroozi_rockstar_2013}. Briefly, Rockstar identifies halos by applying an adaptive, hierarchical friends-of-friends algorithm to identify bound, overdense structures, i.e. halos. Masses are associated with these halos by computing the mass within a sphere which encloses a specified density contrast---in this work we use the density contrast given by the spherical collapse model \citep{bryan_statistical_1998}. Concentrations of halos are found by dividing each halo into 50 concentric shells, computing the density profile in those shells, and then fitting an NFW functional form to that profile. Potential and kinetic energies are computed by direct summation over the constituent particles of each halo.

\begin{figure*}                                                         
 \begin{tabular}{cc}
  \includegraphics[width=85mm]{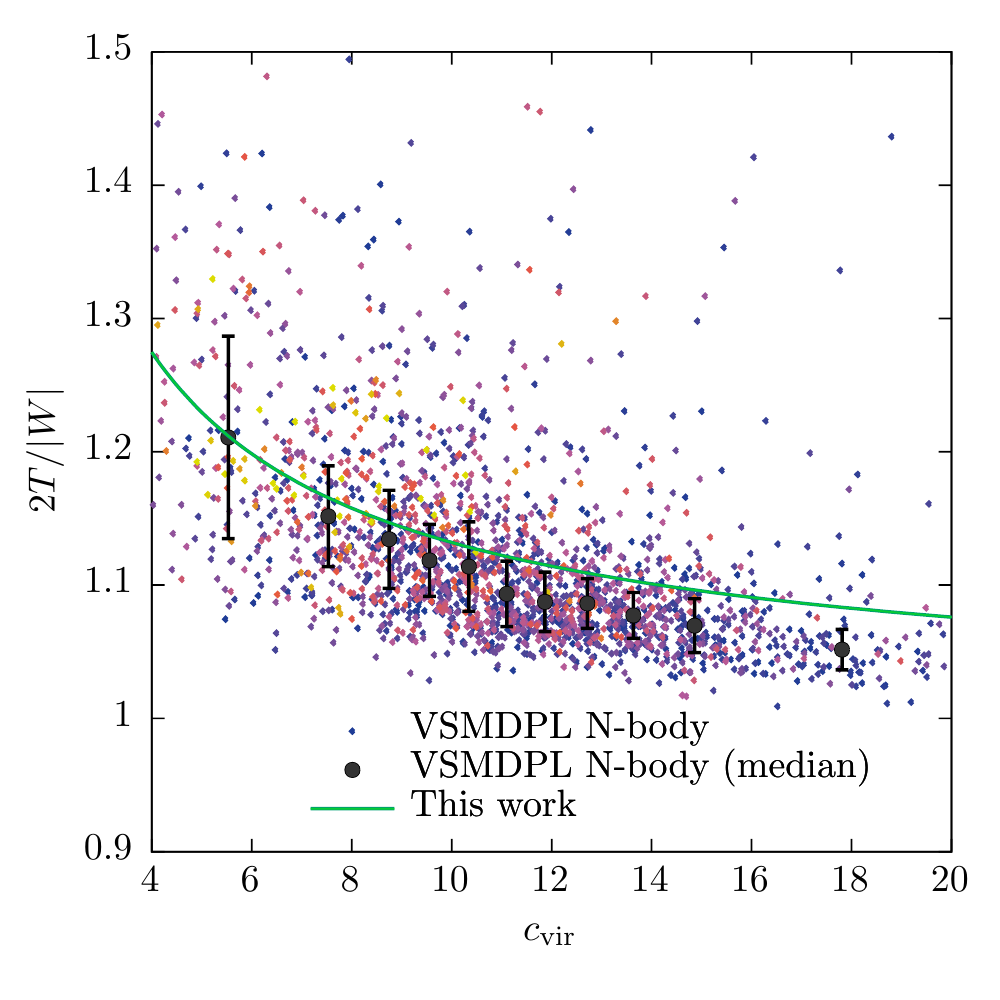} & \includegraphics[width=85mm]{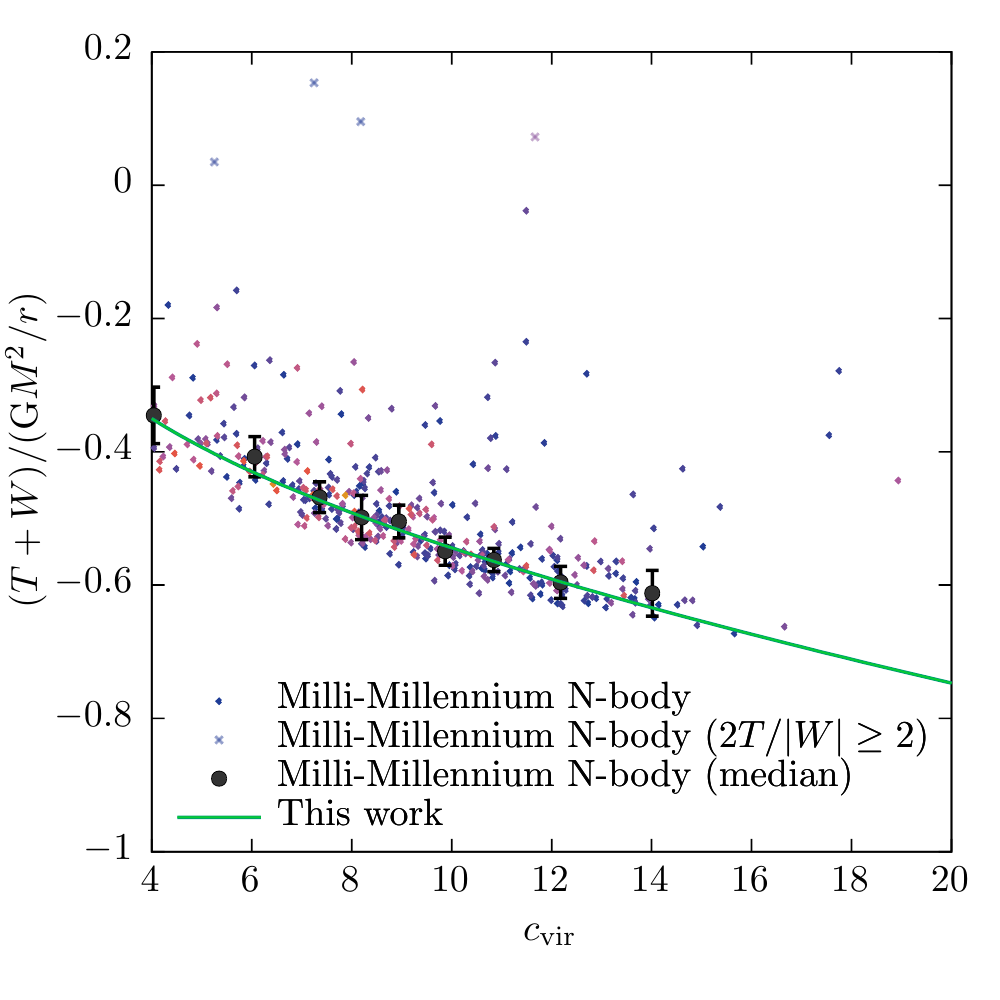}
 \end{tabular}
 \caption{\emph{Left panel:} The virial ratio, $2T/|W|$, as a function of concentration. Multi-colour points show results for individual halos taken from the $z=0$ snapshot of the VSMDPL simulation. All halos containing more than 30,000 particles and the masses of which have less than doubled in the past 3.7~Gyr are included---these selections ensure that halos are well-resolved (such that uncertainties in concentration and virial ratio are small), and that halos far from equilibrium are excluded. The colour of each points represents the halo mass (blue through magenta to yellow, for low to high mass). Black points are the median virial ratio as a function of concentration with the error bars showing the median absolute deviation. The green line is the expectation for the virial ratio for NFW halos computed using equations~(\protect\ref{eq:energyKinetic}) and (\ref{eq:energySelfGravitational}). \emph{Right panel:} The total energy, $T+W$, as a function of concentration. Points and lines are as in the left panel, except that N-body results are derived from the milli-Millennium simulation in this case, with N-body halos selected to have at least 3,000 particles. In computing the mean relation we exclude halos with $2T/|W| \ge 2$ (which are shown by crosses in the figure) as they are far from equilibrium.}
 \label{fig:virialRatio}
\end{figure*}

In the left panel of Figure~\ref{fig:virialRatio} we show the virial ratio as a function of halo concentration. Multi-coloured points show results for individual halos taken from the $z=0$ snapshot of the VSMDPL simulation. All halos containing more than 30,000 particles and the masses of which have less than doubled in the past 3.7~Gyr are included---these selections ensure that halos are well-resolved (such that uncertainties in concentration and virial ratio are small), and that halos far from equilibrium are excluded (as will be discussed below). The colour of each points represents the halo mass (blue through magenta to yellow, for low to high mass). Black points are the median virial ratio as a function of concentration with the error bars showing the median absolute deviation. The green line is the expectation for the virial ratio for NFW halos computed using equations~(\protect\ref{eq:energyKinetic}) and (\ref{eq:energySelfGravitational}). 

The \cite{cole_hierarchical_2000} model for kinetic and potential energies for NFW halos predicts a virial ratio above unity, and varying from over 1.2 for the least concentrated halos, to slightly less than 1.1 for high concentrations, consistent with the findings of \cite{cole_structure_1996}. The \cite{cole_hierarchical_2000} model agrees reasonably well with the median virial ratio found for N-body halos (which also show no strong dependence of virial ratio on halo mass), which follow the same trend with concentration. At high concentrations our model moderately overestimates the median virial ratio, by around 2\%. We conclude that the \cite{cole_hierarchical_2000} model is a good description of the energies of cosmological halos.

While there is considerable scatter in the N-body virial ratio at fixed concentration this does not invalidate our model, which posits a one-to-one mapping between the total energy, $T+W$, and concentration, not between the virial ratio and concentration. The source of this scatter likely originates in the fact that cosmological halos are not the idealized spherically-symmetric, smooth objects that we model here, but have significant triaxiality and substructure. nor are they truly isolated, virialized systems.

The virial ratio is also expected to oscillate around its equilibrium value of approximately $1$ as halos are constantly being perturbed by mergers and interactions with nearby halos. Such oscillations are expected to damp away after a few crossing timescales \citep{lynden-bell_statistical_1967}, but are constantly being re-excited by new mergers and interactions. Nevertheless, the total energy of the system is conserved throughout these oscillations.

We can also compare the expectations of the \cite{cole_hierarchical_2000} model to the \emph{total} energy, $T+W$, of N-body halos. While this is unavailable for the VSMDPL simulation, it can be measured from the milli-Millennium Simulation (a smaller volume simulation otherwise identical to the Millennium Simulation; \citealt{springel_simulations_2005}). To do so, we use the Rockstar halo finder \citep{behroozi_rockstar_2013} to identify halos and measure their properties (mass, concentration, potential and kinetic energy), as described above. As can be seen from equations~(\ref{eq:energyPotential}) and (\ref{eq:energyKinetic}) the total energy of a halo is expected to scale as $\mathrm{G}M^2/r$. We therefore normalize energies by this factor. The right panel of Figure~\ref{fig:virialRatio} shows the resulting relation between total energy and concentration. Once again, colored points indicate individual halos from the milli-Millennium simulation---we select halos with at least 3,000 particles to ensure that they are well resolved, and exclude any halo with $2T/|W|\ge 2$ to avoid including halos which are far from virial equilibrium. (Halos with $2T/|W|\ge 2$ are shown as faint crosses in the figure, and represent only 1.4\% of the total number of halos.) The large blue points show the median of the N-body relation in several bins in concentration. A clear correlation between total energy and concentration is apparent---validating the fundamental assumption of this work. While there is some scatter in total energy at fixed concentration this is relatively small (around 0.05~dex). We note that if the small number of out-of-equilibrium halos (those with $2T/|W|\ge 2$) were included when computing the mean, the results would be shifted by less than 2\%, and so have negligible effect on our interpretation of these results.

The line in the right panel of Figure~\ref{fig:virialRatio} shows the expectation from the \cite{cole_hierarchical_2000} model for NFW halos. It agrees very closely with the N-body halo median, matching the trend with concentration.

At a fixed energy of $-0.5$ (in these dimensionless units) the scatter in concentration is around 0.06~dex. While any scatter in this relation will limit the ability of our model to predict concentration from total energy, the scatter measured here is small. For comparison, \protect\cite{ludlow_mass-concentration-redshift_2016} used their model to predict the concentrations of individual halos from their merger histories, finding residuals of around 0.09~dex for cold dark matter halos at $z=0$ (see their Figure~10 and discussion in their Section~4.4).

Given these results, we expect our model to provide a reasonably accurate description of halo concentrations, while noting that the presence of scatter in the total energy at fixed concentration means that our model is, of course, approximate.

From equations~(\ref{eq:energyPotential}) and (\ref{eq:energyKinetic}) we can see that the total energy, $E=T+W$, is a function of the two parameters $\rho_\mathrm{s}$ and $r_\mathrm{s}$ of the \cite{navarro_universal_1997} profile. Equivalently, we can write $E(M,r_\mathrm{s})$ where $M=M(r_\mathrm{v})$ since the density normalization can be derived from the mass, scale radius, and the choice of density contrast used to define halo mass.

Therefore, the energy of a halo of known mass depends only on $r_\mathrm{s}$---so if we know the energy of the halo the scale radius can be determined. We will next describe how we compute the total energy, $E$, for each halo in a merger tree.

This is the key assumption of our model---the concentration of a halo of known mass can be inferred uniquely from the total energy of the halo. As described above, this assumption relies on the assumptions of spherical symmetry, Jeans equilibrium, and NFW halo profiles. In reality, and in N-body simulations, these assumptions will all be violated to some degree, breaking the one-to-one correspondence between energy and concentration which underpins our model. The utility of our model in the face of these issues will be judged by its ability to match the distribution of concentrations found in N-body simulations, and to predict correlations between those concentrations and other halo properties.

\subsection{Algorithm for scale radii}

Beginning with a merger tree we perform a depth-first walk of the tree, visiting each halo in turn. In what follows, we refer to any progenitor halos of the halo currently being visited using subscript ``$i$'', with $M_i$ being the virial mass of that progenitor halo, and $r_i$ its virial radius. The total energy of each halo is then computed as follows.

\textit{Halos with no resolved progenitors:} For these halos we determine a scale radius by assigning a concentration from the mass-concentration-redshift relation proposed by \cite{diemer_universal_2015}. The energy of the halo is then computed using equations~(\ref{eq:energyPotential}) and (\ref{eq:energyKinetic}). We have checked that our results for scale radii are insensitive to this choice. For example, if we instead assign a fixed concentration of $c=10$ to halos with no resolved progenitor (such that there is no concentration-mass relation imposed on these halos), our results for $z=0$ halos are not significantly affected. This implies that, for a sufficiently well-resolved merger history, the final scale radius of a halo is insensitive to the scale radii of its distant progenitors.

\textit{Halos with one or more progenitors:} For halos with one or more resolved progenitor halos, we label progenitors by an index $i=1$ to $N$, with $i=1$ corresponding to the most massive progenitor. We set the energy equal to the sum of the energies of progenitors and the contribution from unresolved accretion:
\begin{equation}
    E = \left[\sum_{i=1}^N E_{\mathrm{p},i} (1+\mu_i)^{-\gamma} + E_\mathrm{u}\right](1+b),
    \label{eq:energyTotal}
\end{equation}
where $E_{\mathrm{p},i}$ is the energy of the $i^\mathrm{th}$ progenitor, $\mu_i = M_i/M_1$, $E_\mathrm{u}$ is the energy of unresolved accretion, and $b$ and $\gamma$ are parameters whose values we will determine by requiring our model to match N-body simulation results. The combination $(b,\gamma)=(0,0)$ would correspond to the ``ideal'' case in which energy is precisely conserved. However, as we will show in \S\ref{sec:results}, this ideal model does not match the N-body results.

The terms in equation~(\ref{eq:energyTotal}) containing the parameters $b$ and $\gamma$ are not physically motivated, instead being empirical modifiers. They are introduced to account for the approximations made by our model (such as spherical symmetry, Jeans equilibrium, etc.) and to allow it to be calibrated to match results from N-body simulations. We expect each parameter to be of order unity. Specifically, $b$ acts as an overall ``boost'' factor by which the energy of a halo exceeds that of its progenitors, while $\gamma$ acts to create an additional dependence of the energy contributed by each progenitor on the mass of that progenitor.

For resolved progenitors we can write $E_{\mathrm{p},i} = E_{\mathrm{p,int},i} + E_{\mathrm{p,orb},i}$, where $E_{\mathrm{p,int},i}$ is the internal energy of the progenitor (which will have already been computed by virtue of the nature of our depth-first tree walk), and $E_{\mathrm{p,orb},i}$ is the orbital energy of the progenitor. To find the orbital energy, for $i>1$ we select an orbit for each progenitor from the distribution proposed by \cite{jiang_orbital_2015} and compute
\begin{equation}
    E_{\mathrm{p,orb},i} = \left[-{\mathrm{G} M_1 M_i \over r_1} + {1\over 2} {M_i v_i^2 \over 1+\mu_i}\right] ,
    \label{eq:energyOrbital}
\end{equation}
where $v_i$ is the orbital speed of the merging progenitor halo. For the most massive progenitor, $i=1$, we assume no orbital energy (as this progenitor defines the primary ``branch'' of the tree into which the other progenitors are merging). Note that, unlike in \cite{benson_random-walk_2020}, where we were concerned about a vector quantity (angular momentum), here we do not need to consider correlations between the orbital angular momenta of infalling satellite halos and that of the primary halo, as this does not affect the energy.

The mass of material accreted in unresolved halos is taken to be $M_\mathrm{u} = M_\mathrm{d} - \sum_{i=0}^N M_\mathrm{i}$, where $M_\mathrm{d}$ is the mass of the halo whose energy we are computing. We then set $E_\mathrm{u} = \beta M_\mathrm{u} (\epsilon_\mathrm{orbit} + \epsilon_\mathrm{internal})$ where $\epsilon_\mathrm{orbit}$ and $\epsilon_\mathrm{internal}$ are given by equations~(\ref{eq:energyUnresolvedOrbital}) and (\ref{eq:energyUnresolvedInternal}) respectively, and $\beta$ is determined in Appendix~\ref{app:unresolvedResStudy}. Note that $\beta$ (not to be confused with the parameter $b$ introduced above) is not a free parameter of our model, but is instead fixed by requiring that our algorithm for scale radii be stable with respect to the mass resolution of the merger trees used.

After completing this tree walk, each halo with one or more progenitors in the tree has been assigned a total energy, $E$, from equation~(\ref{eq:energyTotal}). We then step back through the tree and compute the corresponding scale radius such that $E(M,r_\mathrm{s})=E$.

\subsection{Calibration}

Our model has two free parameters, $b$ and $\gamma$. To calibrate these parameters we make use of the distributions of concentration parameter measured from the COCO N-body simulations \citep{hellwing_copernicus_2016} by \cite{benson_halo_2019}. To do so we generate a sample of merger trees, and compute scale radii for them using the approach described in the previous subsection. When generating these trees we match the cosmological parameters and power spectrum of the COCO simulation, and sample a total of 40,000 $z=0$ halo masses from the halo mass function between $M_0=10^9\mathrm{M}_\odot$ and $10^{13}\mathrm{M}_\odot$, growing a tree for each sampled mass. 

After building each tree and computing the scale radius of the $z=0$ halo, we determine the concentration parameter $c_\mathrm{200c}=r_\mathrm{200c}/r_\mathrm{s}$, where $r_\mathrm{200c}$ is the radius enclosing a mean density equal to 200 times the critical density, to match the definition used by \cite{benson_halo_2019} when measuring concentrations from the COCO simulations. We then construct distributions of concentrations in seven logarithmically-spaced mass bins spanning the range $9.4 < \log_{10}(M/\mathrm{M}_\odot) < 12.4$, matched to the same mass intervals in which distributions of concentrations were measured from the COCO simulations. Within each mass range, we construct a binned histogram of concentrations which is then normalized to provide an estimate of the distribution of concentrations within the mass bin. That is, we compute
\begin{equation}
p_i \equiv {\mathrm{d}P_i \over \mathrm{d}\log_{10}c_\mathrm{200c}} = {N_i \over \Delta\log_{10} c_\mathrm{200c} \sum_j N_j},
\label{eq:distributionEstimator}
\end{equation}
where $N_i$ is the number of halos with concentrations falling within the $i^\mathrm{th}$ bin of concentration, $\Delta\log_{10} c_\mathrm{200c}$ is the width of each concentration bin, and the sum in the denominator is taken over all concentration bins. As in \cite{benson_halo_2019} we smooth the resulting distributions by a Gaussian with width chosen to match the expected uncertainty in N-body halo concentration estimates \citep[][ equation~5]{benson_halo_2019}. We use the exact same approach to estimate concentration distributions from the COCO simulation (except that there is no need to for the final smoothing step).

Finally, we define the likelihood of our model given the N-body data in each mass interval as:
\begin{equation}
\log\mathcal{L} = -\frac{1}{2} \Delta \mathbf{C}^{-1} \Delta^{\rm T},
\end{equation}
where $\Delta$ is a vector of differences between the N-body and model concentration histograms, and $\mathbf{C}$ is a covariance matrix. The covariance matrix is taken to be the sum of that of the N-body halo histogram, and that of the model halo histogram\footnote{The model covariance can be made arbitrarily small by increasing the number of merger trees that we generate, at the expense of increased computation time. We choose the number of trees generated such that the sample variance is smaller than, but comparable to, the sample variance of the N-body dataset. In this way, our results are not strongly affected by the model sample variance, while we keep computation time within reasonable limits.}, $\mathbf{C}=\mathbf{C}_\mathrm{N-body}+\mathbf{C}_\mathrm{model}$. We assume that the uncertainty in the number of halos contributing to each bin is $\sqrt{N_i}$,
i.e. we assume Poisson statistics. These uncertainties, for both the N-body and model concentration distributions, are then used to construct the respective covariance matrices, including the off-diagonal terms which arise from the fact that the $N_i$ appear in the denominator of equation~(\ref{eq:distributionEstimator}). Specifically,
\begin{equation}
  \mathbf{C} =  \mathbf{J} \mathbf{C}^\prime \mathbf{J}^\mathrm{T}  
\end{equation}
where $\mathbf{J}$ is the Jacobian matrix:
\begin{equation}
  J_{ij} = {\mathrm{d} p_i \over \mathrm{d} N_j},
\end{equation}
and 
\begin{equation}
  C^\prime_{ij} = \left\{ \begin{array}{ll} N_i^2 & \hbox{if } i=j \\ 0 & \hbox{otherwise.} \end{array} \right.
\end{equation}
Additionally, the Gaussian smoothing that was applied to the model histograms is accounted for when computing the covariance matrix for our model.

When constructing the histogram of model halo concentrations we exclude halos which have more than doubled their mass in the last 1.25 crossing times \citep[3.7~Gyr;][ \S2.1.1]{ludlow_mass-concentration-redshift_2016}. This selection criterion was applied to the COCO N-body halos from which concentrations were measured by \cite{benson_halo_2019} to remove out of equilibrium halos (for which concentration may not be well defined), and so we apply the same selection to halos in our model.

We then run an MCMC simulation to determine the posterior distribution over the parameters $b$ and $\gamma$. We follow the same approach in our MCMC methodology as \cite{benson_random-walk_2020}, utilizing 64 parallel chains, and generating proposals using differences between chains. We allow the simulation to progress until all chains are converged. We judge convergence using the Gelman-Rubin statistic, $\hat{R}$ \citep{gelman_a._inference_1992}, after removing outlier chains (identified using the Grubb's outlier test \citep{grubbs_procedures_1969,stefansky_rejecting_1972} with significance level $\alpha=0.05$). Convergence is assumed once $\hat{R}$ reaches a value of $1.2$ in the parameters of interest, $b$ and $\gamma$.

In addition to $b$ and $\gamma$, we include several nuisance parameters in our MCMC simulation, which we will marginalize over in the final analysis. All parameters are described below. For $b$ and $\gamma$ we also detail the priors used, while for nuisance parameters we adopt the same priors as in \cite{benson_random-walk_2020}.
\begin{itemize}
 \item $b$: Energies of merging subhalos are multiplied by $1+b$ (see equation~\ref{eq:energyTotal}). This parameter is therefore expected to be of order unity. We adopt a uniform prior for $b$ between $-1$ and $+1$.
 
 \item $\gamma$: This parameter appears as the exponent of the mass-dependent correction to the energy of merging halos (see equations~\ref{eq:energyTotal} and \ref{eq:energyOrbitalUnresolved}). We adopt a uniform prior between $-2$ and $+2$.

 \item $(A,a,p)$ in the \cite{sheth_ellipsoidal_2001} mass function.
 
 \item $(G_0,\gamma_1,\gamma_2)$ in the halo merger rate model of \cite{parkinson_generating_2008}.

 \item $(b,\gamma,\sigma,\mu)$ in each primary halo mass, and secondary-to-primary mass ratio range in the fitting function for orbital parameters of subhalos of \cite{jiang_orbital_2015}.
\end{itemize}

This gives a total of 44 parameters, although as noted, all except $b$ and $\gamma$ have narrow priors and are included only as nuisance parameters allowing us to marginalize over their uncertainties.

\section{Results}\label{sec:results}

Our MCMC simulation reaches convergence after 1,250 steps. We allow it to run for a further 1,116 steps, and find a correlation length in each chain of around 25 steps. Therefore, our post-convergence chains provide approximately 2,800 independent samples from the posterior distribution over our parameters. The full posterior distribution is shown in Fig.~\ref{fig:modelPosterior}, and we find $ \gamma$ $=1.518^{+0.036}_{-0.036}$ and $ b$ $=(6.730^{+0.049}_{-0.247}) \times 10^{-1}$ when marginalized over all other parameters.

\begin{figure*}
\begin{tabular}{cc}
 \includegraphics[width=85mm]{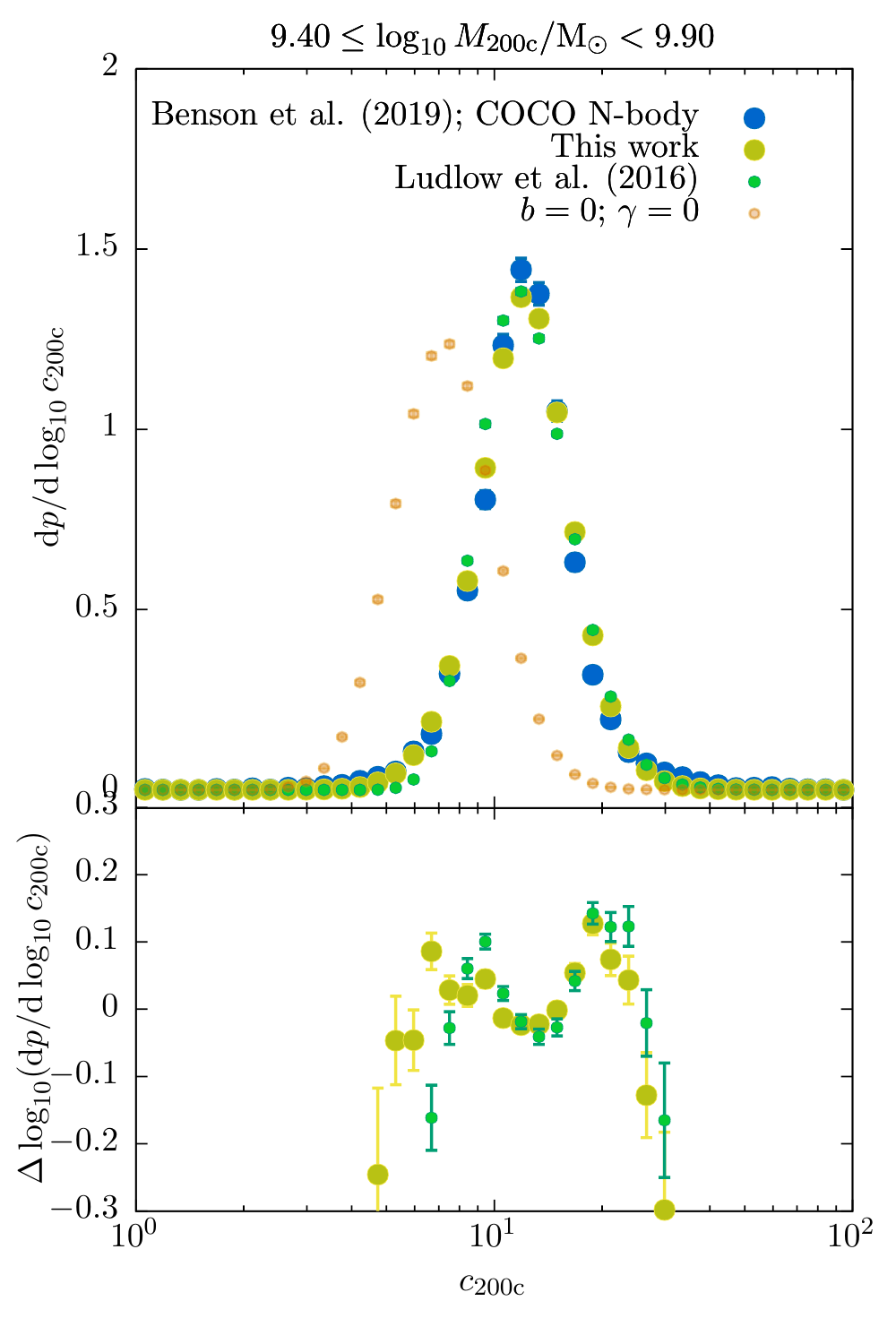} &  \includegraphics[width=85mm]{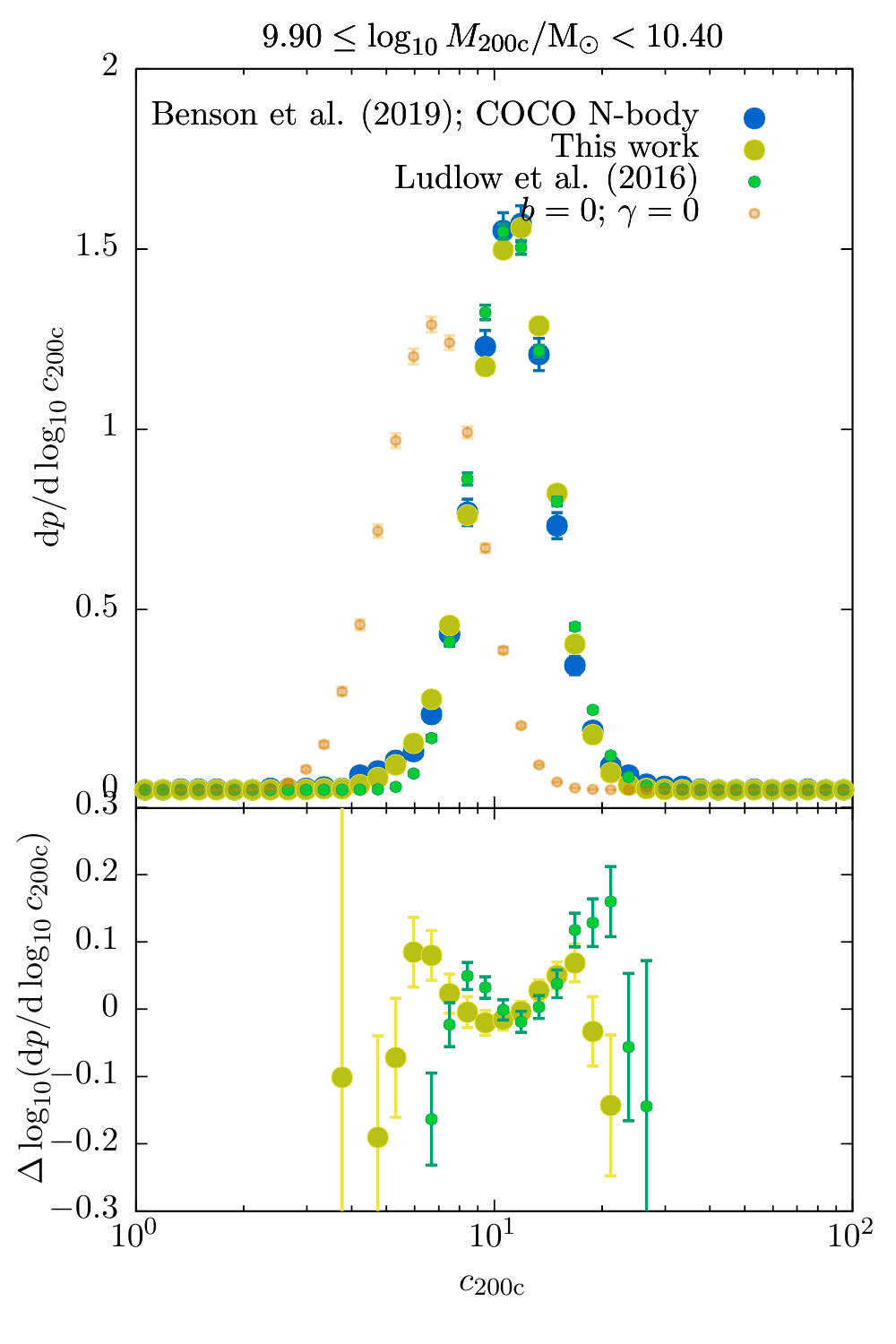}
 \end{tabular}
 \caption{\emph{Upper panels:} Distributions of concentration parameter for $z=0$ halos in two mass ranges (as indicated above each panel). Blue points show the measured distribution from the COCO N-body simulation from \protect\cite{benson_halo_2019}, while yellowish-green points show results from the maximum likelihood model in this work, and smaller green points show results using the model of \protect\cite{ludlow_mass-concentration-redshift_2016} (using the best-fit parameter values from \protect\citealt{benson_halo_2019}). For reference, we show using faint, orange points the distribution obtained from our model by setting $b=0$ and $\gamma=0$ in equation \protect\ref{eq:energyTotal}. Error bars indicate the uncertainty in the results and are estimated assuming that the number of halos falling in each bin of the distribution follows Poisson statistics. In many bins the error bars are smaller than the symbols. \emph{Lower panels:} The logarithmic residuals between the models of this work and of \protect\cite{ludlow_mass-concentration-redshift_2016} and the COCO N-body simulation results, with error bars indicating the uncertainty on this quantity as derived from the error bars in the upper panels. In the tails of the distributions $\Delta\log_{10}(\mathrm{d}p/\mathrm{d}\log_{10}c_\mathrm{200c}) \ll 0$---we show only a limited range on the $y$-axis to emphasize the behavior around the peak of the distribution. Note that the $b=0$, $\gamma=0$ model is not shown in these panels.}
 \label{fig:distribution}
\end{figure*}

The upper panels of Figure~\ref{fig:distribution} shows the distribution of concentrations in two intervals of $z=0$ halo mass from our model (yellowish-green points) and from the COCO simulations (blue points), showing an overall very good agreement. For reference we also show, using faint orange points, the results from our model obtained by setting $b=0$ and $\gamma=0$ in equation \protect\ref{eq:energyTotal}. As can be seen, this ``ideal'' model significantly underpredicts the mean concentration, although it does result in approximately the correct scatter. The lower panels of figure~\ref{fig:distribution} show the logarithmic difference between our model and the COCO simulation results. Note that in the tails of the distributions $\Delta\log_{10}(\mathrm{d}p/\mathrm{d}\log_{10}c_\mathrm{200c}) \ll 0$---we show only a limited range on the $y$-axis to emphasize the behavior around the peak of the distribution. The ``double-peaked'' nature of this logarithmic difference indicates that our model predicts a distribution of concentrations which is not as sharply peaked as the N-body results, while the rapid drop in the logarithmic difference that occurs for low and high concentrations indicates that our model predicts significantly less weight in the tails of the distribution than the N-body results show.

\begin{figure}
\includegraphics[width=85mm]{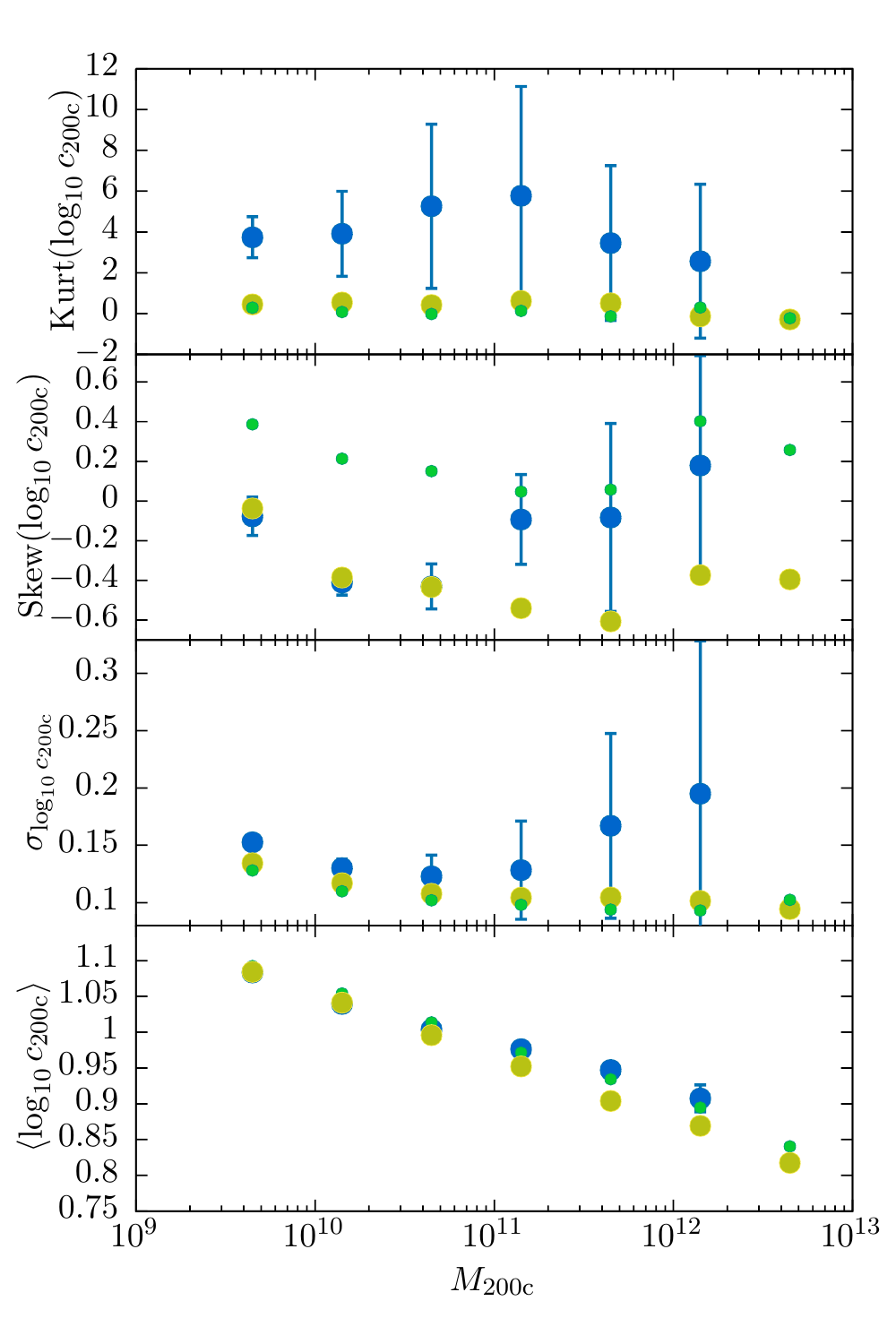}
\caption{Moments of the distribution of concentrations, under the assumption of NFW halo profiles. (For an equivalent figure under the assumption of Einasto halo profiles, see Figure~\protect\ref{fig:statsEinasto}.) Each panel shows a different moment (mean, scatter (i.e. root-variance), skewness, and (excess) kurtosis from bottom to top) as a function of $z=0$ halo mass, with results from this work shown as yellowish-green points, and from the COCO N-body measurements from \protect\cite{benson_halo_2019} as blue points. Also shown, as smaller green points, are results from the model of \protect\cite{ludlow_mass-concentration-redshift_2016} (using the best-fit parameter values from \protect\citealt{benson_halo_2019}). For the N-body results uncertainties arising from the finite number of halos available are found via bootstrapping from the original measurements. Uncertainties due to the finite number of halos used in this work are much smaller and so are not shown.}
\label{fig:stats}
\end{figure}

\begin{table*}
\caption{Moments of the distribution of concentrations, under the assumption of NFW halo profiles. Each pair of rows shows results for a different $z=0$ halo mass range, with results from this work shown on a white background and from the COCO N-body measurements from \protect\cite{benson_halo_2019} on a grey background. The first column indicates the halo mass range, while the remaining columns show the mean, scatter (i.e. root-variance), skewness, and (excess) kurtosis. For the N-body results uncertainties arising from the finite number of halos available are found via bootstrapping from the original measurements. Uncertainties due to the finite number of halos used in this work are much smaller and so are not shown. We indicate with ``--'' cases where the N-body data had too few halos to allow a robust measurement to be made.}
\label{tb:stats}
\begin{tabular}{lllll}
\hline
 & \multicolumn{4}{c}{\textbf{Moments of $\log_{10} c_\mathrm{200c}$}}\\
\textbf{Halo mass} & \textbf{Mean} & \textbf{Scatter} & \textbf{Skewness} & \textbf{Kurtosis}\\
\hline
$ 9.40 \le \log_{10} M_\mathrm{200c}/\mathrm{M}_\odot <  9.90$ & $ 1.08437$ & $ 0.13433$ & $-0.03770$ & $ 0.44587$ \\
\rowcolor{lightgray}    & $ 1.08366 \pm  0.00042$ & $ 0.15258 \pm  0.00600$ & $-0.07742 \pm  0.09698$ & $+3.74604 \pm  0.99892$ \\
$ 9.90 \le \log_{10} M_\mathrm{200c}/\mathrm{M}_\odot < 10.40$ & $ 1.04137$ & $ 0.11687$ & $-0.38536$ & $ 0.54178$ \\
\rowcolor{lightgray}    & $ 1.03958 \pm  0.00043$ & $ 0.13007 \pm  0.00768$ & $-0.41040 \pm  0.06405$ & $+3.90800 \pm  2.08336$ \\
$10.40 \le \log_{10} M_\mathrm{200c}/\mathrm{M}_\odot < 10.90$ & $ 0.99589$ & $ 0.10762$ & $-0.43343$ & $ 0.41786$ \\
\rowcolor{lightgray}    & $ 1.00375 \pm  0.00051$ & $ 0.12272 \pm  0.01842$ & $-0.43141 \pm  0.11352$ & $+5.22451 \pm  4.01680$ \\
$10.90 \le \log_{10} M_\mathrm{200c}/\mathrm{M}_\odot < 11.40$ & $ 0.95217$ & $ 0.10428$ & $-0.54034$ & $ 0.61359$ \\
\rowcolor{lightgray}    & $ 0.97634 \pm  0.00082$ & $ 0.12813 \pm  0.04245$ & $-0.09612 \pm  0.22848$ & $+5.79007 \pm  5.35354$ \\
$11.40 \le \log_{10} M_\mathrm{200c}/\mathrm{M}_\odot < 11.90$ & $ 0.90402$ & $ 0.10436$ & $-0.60677$ & $ 0.50302$ \\
\rowcolor{lightgray}    & $ 0.94719 \pm  0.00480$ & $ 0.16496 \pm  0.08039$ & $-0.09540 \pm  0.47306$ & $+3.37100 \pm  3.78467$ \\
$11.90 \le \log_{10} M_\mathrm{200c}/\mathrm{M}_\odot < 12.40$ & $ 0.86916$ & $ 0.10148$ & $-0.37351$ & $-0.13485$ \\
\rowcolor{lightgray}    & $ 0.90750 \pm  0.01905$ & $ 0.19381 \pm  0.13419$ & $+0.17929 \pm  0.55379$ & $+2.62945 \pm  3.91738$ \\
$12.40 \le \log_{10} M_\mathrm{200c}/\mathrm{M}_\odot < 12.90$ & $ 0.81733$ & $ 0.09442$ & $-0.39473$ & $-0.28924$ \\
\rowcolor{lightgray}    & -- & -- & -- & -- \\
\hline
\end{tabular}

\end{table*}

Figure~\ref{fig:stats} and table~\ref{tb:stats} show the first four moments of the concentration distribution in each halo mass interval. The mean is reproduced in all mass intervals to better than 0.04~dex. The scatter is systematically underpredicted by our model, by around 0.02~dex (the underprediction is larger for the higher mass intervals, but for these the uncertainty in the N-body scatter is too large to confirm that these larger discrepancies are real). To examine why, it is informative to look at the next two moments. The N-body distributions have mild negative skewness, which is reasonably well matched by our model. However, the N-body distributions also have significant positive excess kurtosis (i.e. are leptokurtic). While our model also produces leptokurtic concentration distributions they are not leptokurtic enough compared to the N-body distributions. This can be appreciated by a close inspection of Fig.~\ref{fig:distribution} where the N-body results (blue) points can be seen to have excess in the tails of the distribution compared to the results from our model (yellowish-green points). While the measurements of kurtosis for the N-body data have large uncertainties, the discrepancy from our model is significant.

\subsection{Comparison with Ludlow et al. (2016)}

The model of \cite{ludlow_mass-concentration-redshift_2016} predicts the concentration of a halo utilizing the time at which a given fraction of the halo's mass was first assembled into progenitors above a given mass threshold. \cite{benson_halo_2019} previously calibrated the parameters of the \cite{ludlow_mass-concentration-redshift_2016} model using the same merger trees, and the same N-body calibrator data set as used in this work.

In the $9.40 \le \log_{10} M_\mathrm{200c}/\mathrm{M}_\odot < 9.90$ mass interval \cite{benson_halo_2019} found a mean of $c_\mathrm{200c}=1.077$ and scatter of $\sigma_{\log_{10}c_\mathrm{200c}}=0.114$ utilizing the model of \cite{ludlow_mass-concentration-redshift_2016} together with their model for the dependence of halo merger rates on environment. Our model attains a closer match to both the mean and, significantly, the scatter measured in the N-body simulation than does the \cite{benson_halo_2019} model, although as noted above, our model still does not reproduce the full scatter measured in the N-body halos.

We show the results from the \cite{ludlow_mass-concentration-redshift_2016} model in Figures~\ref{fig:distribution} and \ref{fig:stats} as smaller green points. In Figure~\ref{fig:distribution} it can be seen that the \cite{ludlow_mass-concentration-redshift_2016} model produces results very similar to that of this work---with the notable exception that it significantly underpredicts the tail of N-body halos with low concentrations, while the model described in this work achieves a much better match to this low-concentration tail. This can also be seen in the second panel of Figure~\ref{fig:stats} which shows the skewness in the concentration distribution as a function of halo mass. The \cite{ludlow_mass-concentration-redshift_2016} model significantly overpredicts skewness at low masses, while the model of this work matches the N-body skewness accurately. At higher masses the uncertainty in the N-body skewness becomes too large to discriminate between the models.

Considering the scatter in concentration, our model produces larger scatter (albeit by a small amount) at all masses except for the highest mass bin, and is therefore closer to the N-body results. The model of \cite{ludlow_mass-concentration-redshift_2016} performs somewhat better in matching the mean concentration at higher masses. However, this is largely due to the fact that the \cite{ludlow_mass-concentration-redshift_2016} is positively skewed---the mode of the concentration distribution in the models of \cite{ludlow_mass-concentration-redshift_2016} and of this work are in close agreement across all halo masses.

The origin of the skewness in the distribution of concentrations at fixed mass is unclear. In the \cite{ludlow_mass-concentration-redshift_2016} model this must arise from the structure of the merger trees themselves (as this is the only input to the \cite{ludlow_mass-concentration-redshift_2016} model). The \cite{ludlow_mass-concentration-redshift_2016} model predicts a significant positive skewness, which, as noted above, is due to a lack of halos in the tail of low concentrations. It is possible that these low-concentration halos arise from merger histories which have significant late-time merging, which would increase their energies, making them less bound (and, therefore, less concentrated). Such late-time merger activity may not be captured by the \cite{ludlow_mass-concentration-redshift_2016} model, which considers only the time at which a certain fraction of a halo's final mass was first assembled. In the model presented in this work, such late-time merging \emph{would} affect the halo concentration as we consider the effects of every merging event. 

Of the remaining skewness present in our model it is interesting to ask how much of this arises from the structure of the merger trees, and how much from any skewness in the distribution of merging halo orbital energies. To examine this we ran our model using an artificially modified distribution of merging halo orbital parameters, constructed to have the same mean and variance as the original distribution, but with zero skewness in orbital energy (i.e. we symmetrized the distribution of orbital energies about the mean). We found that this lead to only a small increase in skewness, not significantly changing the level of agreement with the N-body data. The skewness in our model must therefore arise from the structure of the merger trees.

Finally, it is interesting to consider the kurtosis predicted by the \cite{ludlow_mass-concentration-redshift_2016} model. In the top panel of Figure~\ref{fig:stats} we compare the predicted kurtosis to that measured in N-body simulations. As with the model of this work, \cite{ludlow_mass-concentration-redshift_2016} predicts a kurtosis much lower than that measured in the N-body simulations, and, in fact, is largely consistent with the kurtosis predicted by the model of this work. 

This mismatch in kurtosis could indicate that our model (and that of \citealt{ludlow_mass-concentration-redshift_2016}) are failing to capture some of the extremes of halo formation histories, which might indicate some limitation of either the structure of the merger trees used in this work, or in our model for concentrations. Alternatively, it may be that the kurtosis in the distribution of N-body concentrations may be an artefact of how those concentrations are measured. For example, substantial substructure, or non-sphericity in the N-body halos may bias the measurement of concentrations from them. Without greatly improved statistics from the N-body simulations, we avoid drawing any strong conclusions from the differences in kurtosis, and simply note that they indicate some possible, unexplained difference in the tails of the distributions.

\subsection{Tests of the model}

Having constrained our model to match the distribution of $z=0$ halo concentrations as a function of mass, we now explore other statistics predicted by our model and compare them to results from N-body simulations as a way to test the predictive power of our model.

We consider three predictions from our model. In \S\ref{sec:autoCorrelation} we examine the autocorrelation function of halo concentrations across time, in \S\ref{sec:spinCorrelation} we examine the correlation between halo concentration and halo spin parameter, and in \S\ref{sec:higherMass} we examine concentrations of higher mass halos.

For these tests we make use of data from the \href{https://www.cosmosim.org/cms/simulations/vsmdpl/}{VSMDPL} and \href{https://www.cosmosim.org/cms/simulations/bigmdpl/}{BigMDPL} simulations provided via the \href{https://www.cosmosim.org}{CosmoSim} database. From the Rockstar \citep{behroozi_rockstar_2013} halo catalogs derived from the VSMDPL simulation we select $z=0$ halos with masses greater than $2.74 \times 10^{11}\mathrm{M}_\odot$, corresponding to 30,000 particles, and their primary progenitors\footnote{We define ``primary progenitor'' recursively. Let $i$ label snapshots of the simulation at redshifts $z_i$ such that $z_0=0$ and $z_{i+1} > z_i$. The primary progenitor at snapshot $i$, $p_i$, is defined as $p_i=\mathrm{mmp}(p_{i-1})$ where $\mathrm{mmp}(p)$ is a function that selects the most massive progenitor of halo $p$. Note that this means that $p_i$ is not necessarily the most massive halo in a merger tree at snapshot $i$, but instead corresponds to the halo reached by stepping back through the halos of the merger tree, always moving to the most massive progenitor of the current halo.} down to masses of $2.74 \times 10^9\mathrm{M}_\odot$, corresponding to 300 particles. This gives a sample of 77,099 halos which are sufficiently well resolved that we can follow their structure back to early times in the simulation.

We select halos from the BigMDPL simulation in a similar way, but limit our selection to halos with a $z=0$ virial mass of at least $5 \times 10^{14}M_\odot$ (corresponding to over 14,000 particles).

\subsubsection{Auto-correlation function}\label{sec:autoCorrelation}

\begin{figure}
 \includegraphics[width=85mm]{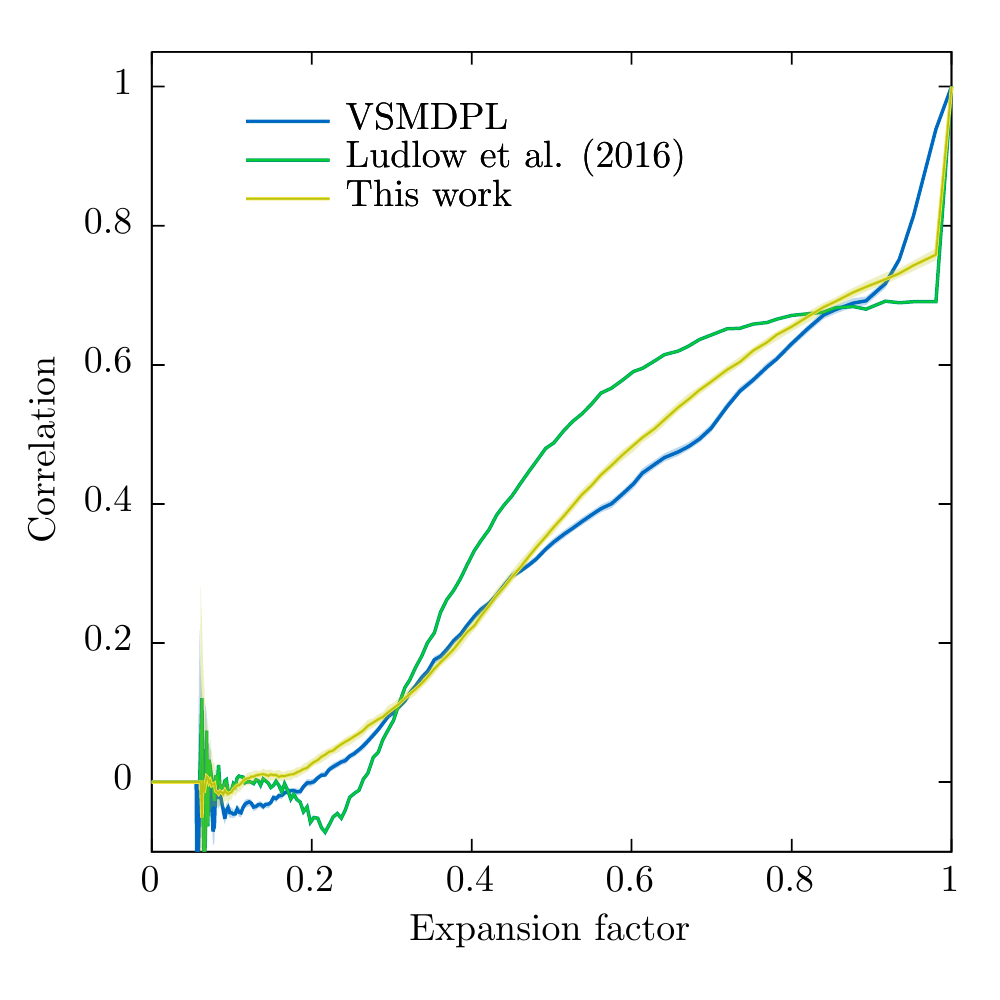} 
 \caption{The autocorrelation of halo scale radius as a function of expansion factor for our sample of $z=0$ halos. The blue line indicates results from the sample of halos extracted from the VSMDPL N-body simulation, with the light blue shaded region indicating bootstrap uncertainties on this quantity. The yellowish-green line and shaded region shows the same results from the maximum likelihood model of this work, while the green line and shaded region shows results from the \protect\cite{ludlow_mass-concentration-redshift_2016} model.}
 \label{fig:autocorrelation}
\end{figure}

In Fig.~\ref{fig:autocorrelation} we show the autocorrelation function of normalized halo scale radius (described below) for this sample of N-body halos from the VSMDPL simulation as a function of expansion factor (blue line, with uncertainties determined from bootstrapping indicated by the blue shading).

Since our sample includes halos with a wide range of scale radii, we normalize scale radii to that expected under the mean concentration-mass relation of \cite{gao_redshift_2008}, and then compute the autocorrelation function of these normalized scale radii. We follow \citeauthor{ludlow_mass-concentration-redshift_2016}~(\citeyear{ludlow_mass-concentration-redshift_2016}; see also \citealt{benson_halo_2019}) and exclude halos which have more than doubled their mass in the past 3.7~Gyr in order to exclude systems which may be out of equilibrium.

Measurements of halo concentration (and, therefore, scale radius) from N-body simulations are affected by uncertainties due to the finite number of particles present in the N-body halo, and to the non-smooth, non-spherical nature of cosmological halos. In Fig.~\ref{fig:autocorrelation} the initial rapid drop in the correlation function (from expansion factor $a=1$ to $a\approx 0.9$) is a consequence of these measurement uncertainties. For $a<0.9$ the continued decline in the correlation function is driven by real, physical decorrelation of halo scale radii. As can be seen from Fig.~\ref{fig:autocorrelation}, the correlation drops to $0.5$ at an expansion factor of $a \approx 0.7$, corresponding to a look-back time of $t \approx 4.6$~Gyr. This can be compared to the timescale for linear growth $D/\dot{D}$ (where $D(t)$ is the linear growth factor, and an overdot represents a time derivative) which is the timescale on which we expect cosmological structures to grow, and which is approximately 7~Gyr at $z=0$. We can conclude that the concentration of a halo changes significantly on a timescale comparable to the timescale of structure growth.

To compute the correlation function using the model from this work we build a set of approximately 36,000 merger trees using parameters sampled from the posterior distribution of our MCMC simulation. We match cosmological and power spectrum parameters to the VSMDPL simulation, and span the same range of masses for $z=0$ halos as for the VSMDPL simulation. The resolution of each merger trees is set to $M_\mathrm{res} = \mathrm{min}(10^{-3} M_0, 2.74 \times 10^9 \mathrm{M}_\odot)$ such that all trees resolve progenitors to at least the mass limit for progenitors used in our correlation analysis, and with at least enough resolution to ensure that our concentrations are well-converged. For these trees, we output the primary progenitor at the set of redshifts corresponding to the snapshots available in the VSMDPL simulation.

As our determinations of scale radii for these trees are not affected by the same uncertainties that affect measurements from N-body simulations, we must add noise to our scale radii to mimic this effect in order to permit a fair comparison with the N-body results. \cite{benson_halo_2019} determined the uncertainty for their measurements of concentration from the COCO N-body simulations and provide a fitting function for the uncertainty, as a function of halo particle number and concentration. We adopt this model here, but allow some freedom in the choice of normalization (the lead term in the expression $a(c_\mathrm{200c})$ in equations~4 \& 5 of \citealt{benson_halo_2019}) as the details of the way in which scale radii were measured by \cite{benson_halo_2019} and by \textsc{Rockstar} (which was used for VSMDPL) differ. While \cite{benson_halo_2019} found a value of $-0.20$ for this term for cosmological halos, we find that a value of $+0.20$ is required to match the uncertainties in the VSMDPL \textsc{Rockstar} halo catalogs.

The results are shown in Fig.~\ref{fig:autocorrelation} by the yellowish-green region, which spans the $10^\mathrm{th}$ and $90^\mathrm{th}$ percentiles of the posterior distribution, with the central line showing the $50^\mathrm{th}$ percentile (i.e. median).

The effects of the mimicked N-body uncertainties are clearly seen in the red line, which drops sharply at the first snapshot at expansion factor $a<1$. (Note that the VSMDPL line drops over a few snapshots, presumably because some of the noise effects that cause this drop are correlated between snapshots---we do not attempt to model that correlation in our mimicked noise). If this noise is removed from our model scale radii the yellowish-green curve instead declines smoothly.

Once these numerical effects are accounted for, our model closely matches the dependence of the correlation function on expansion factor measured from the VSMDPL simulation, indicating that its predicted evolution of scale radii in time is consistent with N-body results.

We also show in Figure~\ref{fig:autocorrelation} results from the \cite{ludlow_mass-concentration-redshift_2016} model (green line and shaded region). At intermediate times this model predicts stronger correlation with the $z=0$ concentration than our model. This should be expected as concentration in the \cite{ludlow_mass-concentration-redshift_2016} model depends on a measure of the formation time of each halo, rather than on the entire merging history. As such, individual merging events in a halo's formation history will have a stronger effect on the concentration in our model than in the \cite{ludlow_mass-concentration-redshift_2016} model.

\subsubsection{Concentration-Spin Correlation}\label{sec:spinCorrelation}

\begin{figure}
 \includegraphics[width=85mm]{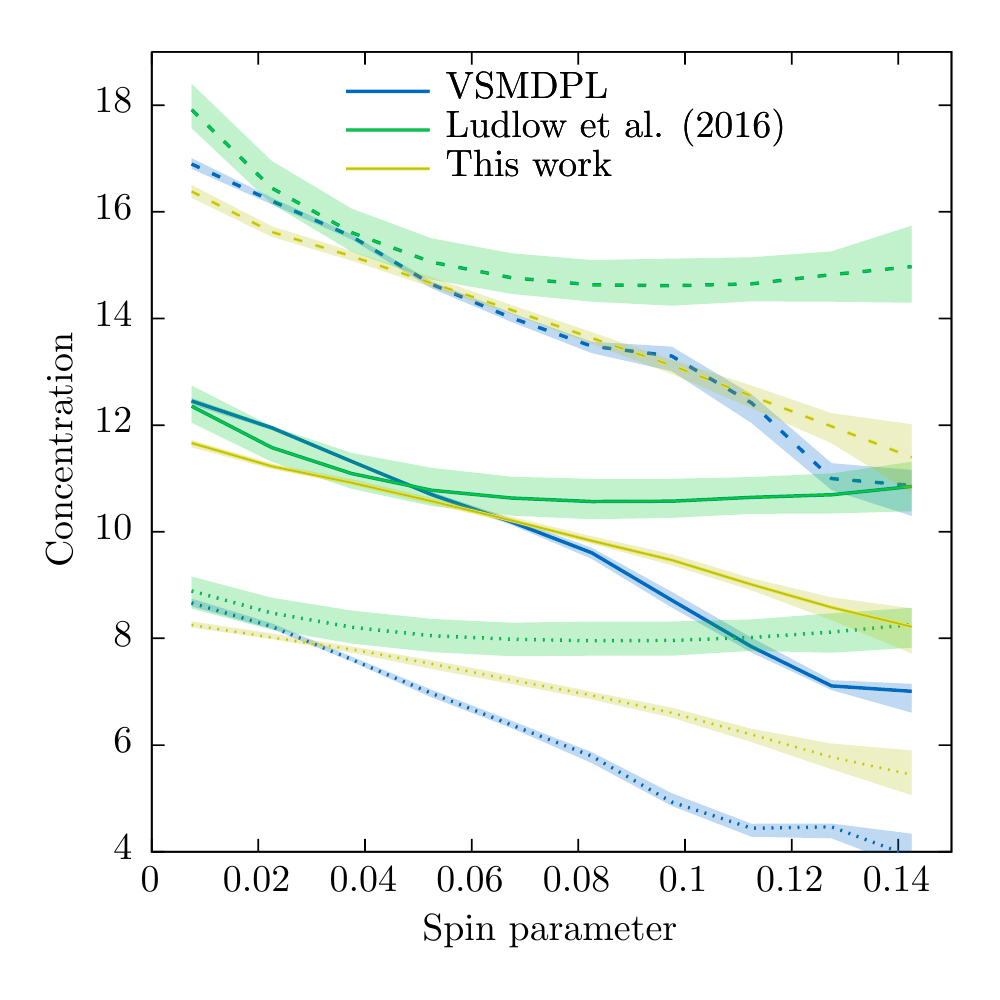} 
 \caption{The $10^\mathrm{th}$, $50^\mathrm{th}$, and $90^\mathrm{th}$ percentiles (dotted, solid, and dashed lines respectively) of the distribution of concentrations as a function of spin parameter for $z=0$ halos selected from the VSMDPL N-body simulation (blue lines), for halos created using the model described in this work, and augmented by the model for halo spin from \protect\cite[][yellowish-green lines]{benson_random-walk_2020}, and for the model of \protect\cite{ludlow_mass-concentration-redshift_2016} also augmented by the model for halo spin from \protect\cite[][green lines]{benson_random-walk_2020}. Shaded bands show the uncertainties in these percentiles. In the case of the VSMDPL simulation uncertainties are found by bootstrap resampling of the N-body halos, while in the case of the results from this work the uncertainties are found by marginalizing over the posterior distribution of the model parameters.}
 \label{fig:concentrationSpinCorrelation}
\end{figure}

We next examine any possible correlation between halo concentration and spin parameter. Figure~\ref{fig:concentrationSpinCorrelation} shows the $10^\mathrm{th}$, $50^\mathrm{th}$, and $90^\mathrm{th}$ percentiles (dotted, solid, and dashed lines respectively) of the distribution of concentrations as a function of spin parameter using the same set of halos from the VSMDPL N-body simulation as used in \S\ref{sec:autoCorrelation} (blue lines), as well as from our matched set of merger trees generated using the model developed in this paper (yellowish-green lines), to which we now additionally apply the random-walk model for halo spins\footnote{Note that we include the correlations between orbital parameters of merging halos and the spin vector of the halo with which they merge as required by the \protect\cite{benson_random-walk_2020} model for spins. This has no effect on the concentration model developed in this work.} developed by \cite{benson_random-walk_2020}, allowing us to simultaneously predict both concentrations and spin parameters. Shaded bands show the uncertainties in these percentiles. In the case of the VSMDPL simulation, uncertainties are found by bootstrap resampling of the N-body halos, while in the case of the results from this work the uncertainties are found by marginalizing over the posterior distribution of the model parameters. For comparison we also include the results of applying the model of \cite{ludlow_mass-concentration-redshift_2016} to these same merger trees (green lines).

We again mimic the numerical uncertainties in concentrations from the N-body simulation as in \S\ref{sec:autoCorrelation}, and now also mimic numerical uncertainties in spins using the model of \cite{benson_constraining_2017}. The \cite{benson_constraining_2017} model for spin noise contains two parts: a spin-independent term which describes the random walk in spin as the angular momenta of individual particles are summed, and a spin-dependent term which is driven by the factors of mass, energy, and radius which appear in the definition of spin. Since both this spin-dependent part and the noise in halo concentrations are driven by the same underlying uncertainty in the total number of particles in the halo, we expect them to correlated. In fact, since the $|E|^{1/2} M^{-5/2}$ term appearing in the definition of halo spin \citep{peebles_origin_1969} scales as $M^{-5/3}$, while concentrations scale as $R\propto M^{1/3}$, we expect an upward fluctuation in $M\propto N$ to increase concentration, but decrease spin. We therefore model these error terms as being anti-correlated.

Figure~\ref{fig:concentrationSpinCorrelation} shows a clear correlation between concentration and spin parameter, with higher spin implying lower concentration. This is understandable in the context of our model for halo concentrations coupled with the spin model of \cite{benson_random-walk_2020}. A high spin halo typically results from the merging of a relatively massive halo with large specific angular momentum. Such a halo also has relatively large (i.e. less negative) energy due to the high kinetic energy associated with that large specific angular momentum and high mass---as such it will tend to increase the specific energy of the halo it merges with, thereby lowering its concentration.

Our model predicts a trend of concentration with spin which matches quite closely that measured from the N-body simulation.

In comparison, the model of \cite{ludlow_mass-concentration-redshift_2016} predicts a much weaker trend of concentration with spin. While the correlation at low spin is quite good, at higher spins the \cite{ludlow_mass-concentration-redshift_2016} fails to capture the decline in concentration found in N-body simulations. This is to be expected as these high spin halos, which result from recent, significant mergers will not \emph{strongly} influence the concentration derived in the \cite{ludlow_mass-concentration-redshift_2016} model, which depends on a measure of the formation time of the halo, and not on individual merger events.

\subsubsection{Higher Mass Halos}\label{sec:higherMass}

Our model has been calibrated against results from N-body simulations for halos in the mass range $9.4 \le \log_{10} M_\mathrm{200c}/\mathrm{M}_\odot < 12.9$. We can explore predictions from our model for halos of significantly higher mass by comparing to results from the BigMDPL simulation. As described above, we extract a sample of $M_\mathrm{vir} > 5\times10^{14}\mathrm{M}_\odot$ halos\footnote{Here, $M_\mathrm{vir}$ is defined as the mass enclosed within a sphere with mean density equal to that predicted by the spherical collapse model for the BigMDPL cosmology \citep{bryan_statistical_1998}.} from the BigMDPL simulation. These halos are resolved with over 14,000 particles, sufficient to allow an accurate measure of their scale radii, and allowing us to trace their progenitors back to early times. We exclude from this sample any halos which have more than doubled their mass in the past 3.7~Gyr, in line with our our analysis of both the COCO halos and the model of this work.

\begin{figure}
 \includegraphics[width=85mm]{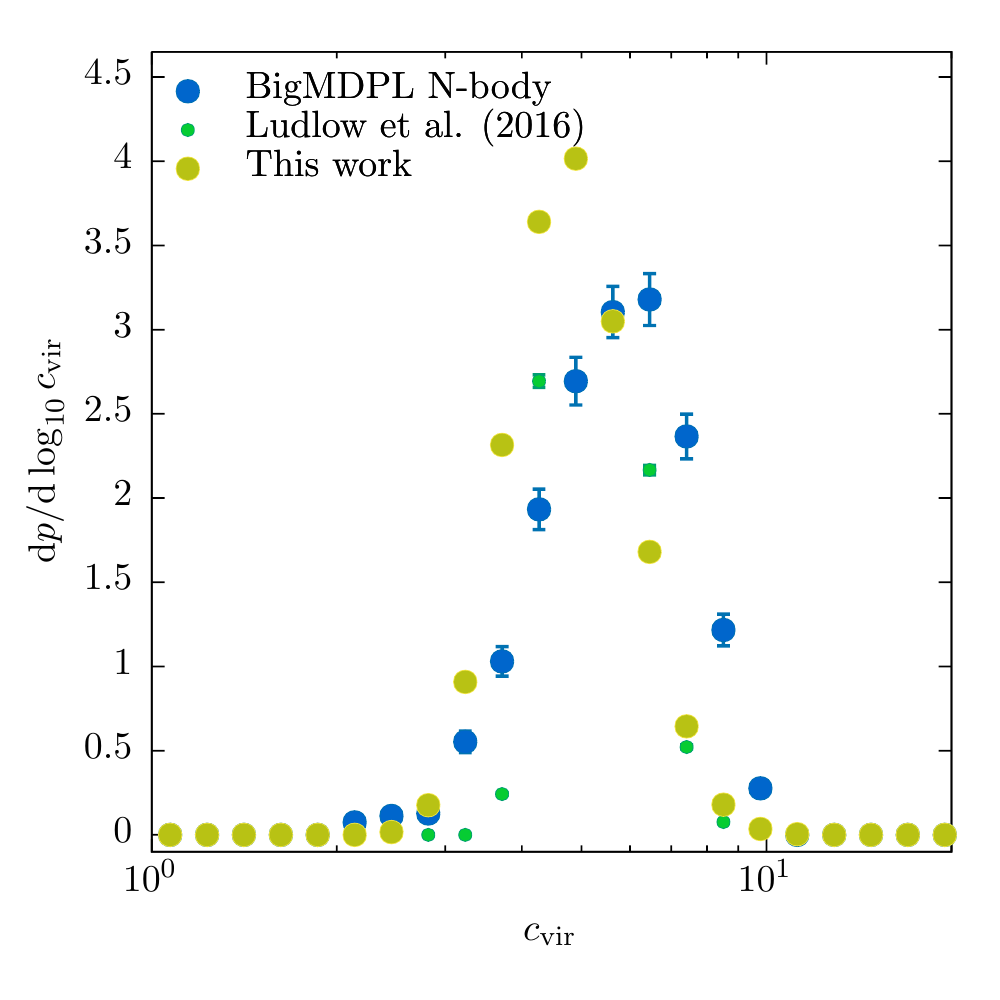} 
 \caption{Distributions of concentration parameter for $z=0$ halos with virial masses $M_\mathrm{vir}>5 \times 10^{14}\mathrm{M}_\odot$. Blue points show the measured distribution from the BigMDPL N-body simulation, yellowish-green points show results from the maximum likelihood model in this work, while green points show the expectations from the model of \protect\cite{ludlow_mass-concentration-redshift_2016}.}
 \label{fig:concentrationsBigMDPL}
\end{figure}

Using the {\sc Rockstar} halo catalogs from BigMDPL we measure the distribution of concentrations, $c_\mathrm{vir}=r_\mathrm{vir}/r_\mathrm{s}$, for these halos. We then generate a set of halos in the same mass range (and using the same cosmological parameters and power spectrum as the BigMDPL simulation) using our model, and measure their concentrations. The results are shown in Figure~\ref{fig:concentrationsBigMDPL}. Concentrations for these higher mass halos are significantly lower, peaking at around $c_\mathrm{vir}=6$ for the BigMDPL halos. Our model predicts a peak at around $c_\mathrm{vir}=5$ ---slightly lower than that found for BigMDPL. The width of the distribution is also larger for BigMDPL. Therefore, while our model does correctly predict significantly lower concentrations for these highest mass halos, it is not in perfect agreement with N-body results. The \cite{ludlow_mass-concentration-redshift_2016} model similarly underpredicts both the mean concentration and the scatter in the distribution for these high mass halos.

\begin{figure}
 \includegraphics[width=85mm]{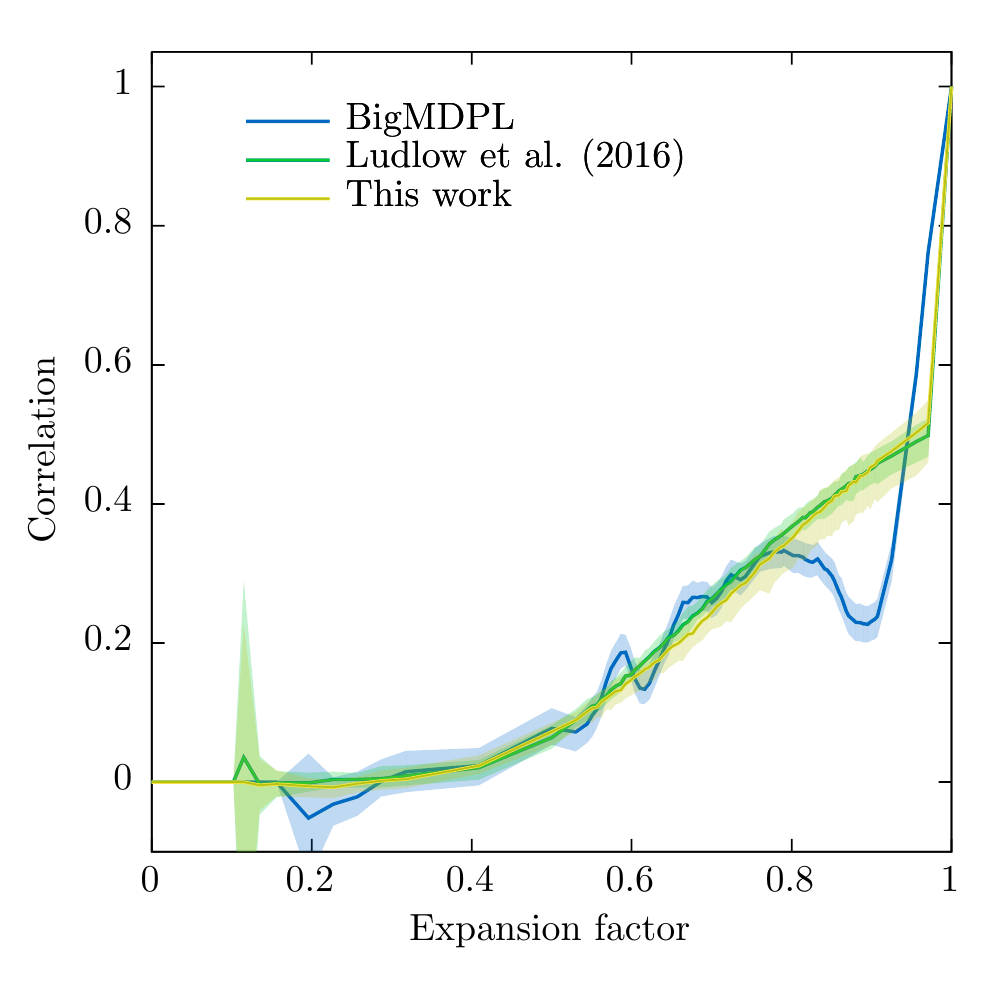} 
 \caption{The autocorrelation of halo scale radius as a function of expansion factor for $z=0$ halos with virial masses $M_\mathrm{vir} > 5\times 10^{14}\mathrm{M}_\odot$. The blue line indicates results from the sample of halos extracted from the BigMDPL N-body simulation, with the light blue shaded region indicating bootstrap uncertainties on this quantity. The yellowish-green shaded region shows the same results for the maximum likelihood model of this work, while the green line and shaded region show the \protect\cite{ludlow_mass-concentration-redshift_2016} model.}
 \label{fig:autocorrelationBigMDPL}
\end{figure}

We also repeat the auto-correlation analysis of \S\ref{sec:autoCorrelation} for our sample of BigMDPL halos, and a match set of halos from our model. Results are shown in Figure~\ref{fig:autocorrelationBigMDPL}. As in the analyis of \S\ref{sec:autoCorrelation}, we see that numerical noise leads to a rapid decorrelation in the N-body halo scale lengths at small look-back times, beyond which the physical correlation can be seen. We again mimic the effect of this numerical decorrelation in the result from our model. The \cite{ludlow_mass-concentration-redshift_2016} model performs equally well in matching the N-body data.

Comparing these results to Figure~\ref{fig:autocorrelation} it can be seen that the scale length in these higher mass halos decorrelates more rapidly than in lower mass halos. This is to be expected since the highest mass halos are still in the ``rapid growth'' stage of their formation. Our model accurately matches the correlation function for these high mass halos, indicating that it correctly captures the effects of rapid growth on the evolution of halo scale radii.

\begin{figure}
 \includegraphics[width=85mm]{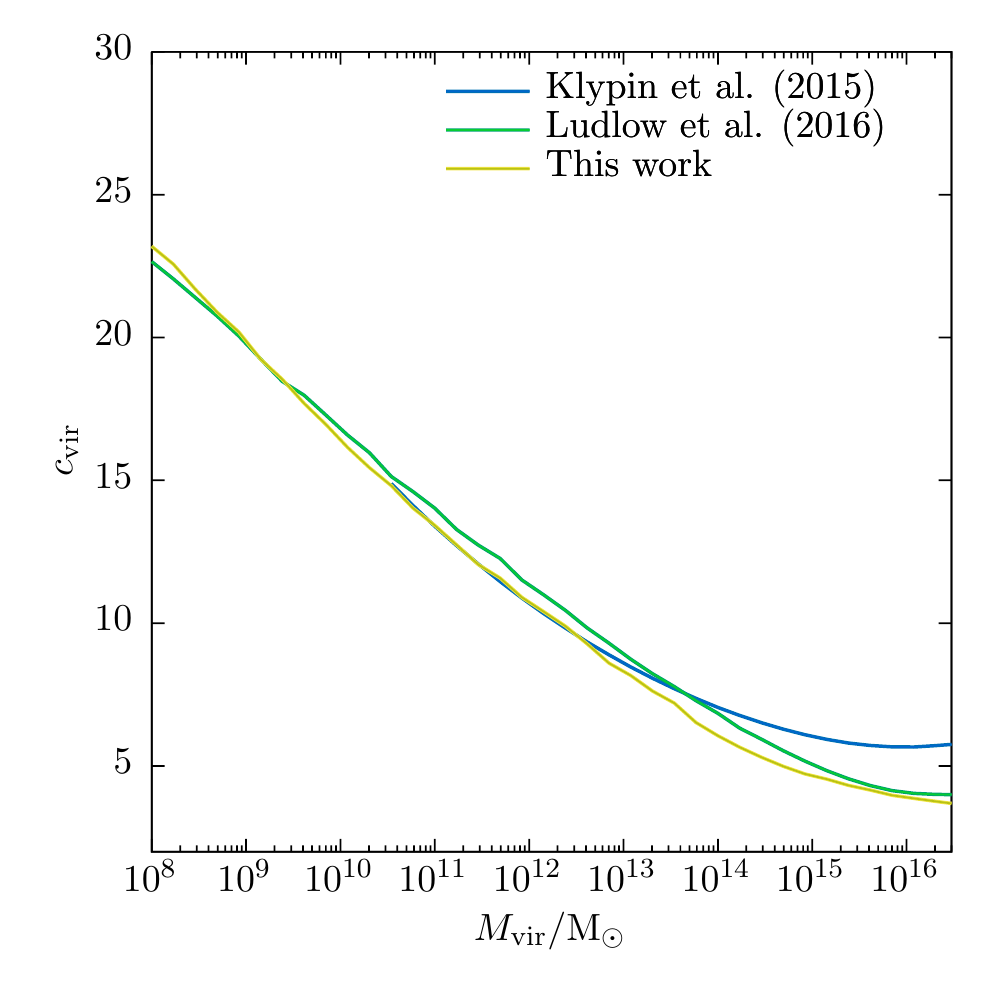} 
 \caption{The mean concentration as a function of halo mass at $z=0$. The blue line shows the model of \protect\cite{klypin_multidark_2016} for the Planck cosmology using in the MultiDark simulations, and for a concentration defined as $c_\mathrm{vir}=r_\mathrm{vir}/r_\mathrm{s}$. The yellowish-green line indicates results from this work for the same cosmological model and concentration definition, while the green line shows results using the \protect\cite{ludlow_mass-concentration-redshift_2016} model. The results for the \protect\cite{klypin_multidark_2016} model are plotted down to the lowest halo mass which was probed by their simulations.}
 \label{fig:trends}
\end{figure}

Finally, in Figure~\ref{fig:trends}, we examine the relation between the mean concentration and halo mass across a very wide range of halo masses, and compare to the results of \cite{klypin_multidark_2016}, who identify a flattening and upturn in the concentration-mass relation at high masses. Our model matches that of \cite{klypin_multidark_2016} well up to halo masses of around $10^{13}\mathrm{M}_\odot$, but then begins to fall below the fit found by \cite{klypin_multidark_2016}. Our model does, however, show a flattening of the relation above $M_\mathrm{vir}\approx 10^{14}\mathrm{M}_\odot$---this is driven in our model by the transition to halos which are in the ``rapid growth'' phase of their assembly. Our model does not produce the upturn in concentration at the highest masses found by \cite{klypin_multidark_2016}. The \cite{ludlow_mass-concentration-redshift_2016} model (shown by the green line in Figure~\ref{fig:trends}) shows very similar behavior to our model, and likewise does not predict an upturn in concentrations at high masses.

\section{Conclusions}\label{sec:conclusions}

We have described a model for predicting the scale radii (or, equivalently, concentrations) of dark matter halos based on their full merging history. The approach is motivated by existing work which demonstrates a clear connection between concentration and halo formation history \cite[e.g.][]{ludlow_mass-concentration-redshift_2016}, but utilizes a random walk model in halo energy inspired by similar treatments for halo angular momentum \citep{vitvitska_origin_2002,benson_random-walk_2020}. Our work therefore has the advantage that it proposes a physical explanation for the causal connection between formation history and concentration. The underlying hypothesis is supported by the existence of a clear correlation, with minimal scatter, between total energy and concentration in N-body halos extracted from cosmological simulations. It is also supported by the work of  \cite{wang_concentrations_2020} who examined the effects of major and minor mergers on halo concentrations, demonstrating that they directly lead to changes in halo structure and scatter in the concentation--mass relation.

While fitting functions or empirical models will undoubtedly be able to provide quantitatively more accurate and precise descriptions of the $c(M,z)$ relation, understanding the physical reasons behind this relation is intrinsically interesting, and may help extend such models to other regimes of mass, power spectrum shape, or dark matter particle properties.

When calibrated, this model closely reproduces the mass-dependent distribution of concentrations measured in N-body simulations. By utilizing the entirety of the information available in a halo's merger tree, our model explains more of the scatter in the concentration-mass relation than previously possible \citep[see, for example,][]{benson_halo_2019}.

Furthermore, we have shown that our model closely matches the auto-correlation function of scale radii across time measured from a high-resolution cosmological N-body simulation, and, when coupled with the random-walk model for halo spins of \cite{benson_random-walk_2020}, also closely matches the correlation between the distribution of concentrations and halo spin parameter measured from that same simulation. In comparison, the \cite{ludlow_mass-concentration-redshift_2016} model performs much less well in matching the N-body results for both of these correlations. We hypothesise that this is because the \cite{ludlow_mass-concentration-redshift_2016} does not follow the effects of individual merger events which can lead to more rapid decorrelation of the concentration over time, and induce correlations between concentration and spin.

Testing our model against higher mass ($M_\mathrm{vir}>5\times 10^{14}\mathrm{M}_\odot$) halos from N-body simulations we find that while it correctly predicts much lower concentrations for these halos it is offset from the peak of the concentration distribution measured from the BigMDPL simulation, and predicts too little scatter in concentration at fixed halo mass. While some of this mismatch might be attributable to the different way in which scale radii were measured from BigMDPL, compared to the analysis of the COCO simulations performed by \cite{benson_halo_2019}, it likely also shows that our model as presently formulated is an imperfect description of these highest mass halos. While our model is currently calibrated using the NFW profile, \cite{klypin_multidark_2016} show that the highest mass halos deviate significantly from the NFW form, being better described by Einasto profiles. As shown in Appendix~\ref{sec:profile} using an Einasto profile would increase the concentrations predicted by our model (although this increase may well be removed if we were to recalibrate our model for the Einasto profile).

Another possible cause of the discrepancy between our model and the high mass halo sample is the assumed distribution of orbital parameters for merging systems. For these we use the model of \cite{jiang_orbital_2015} which is calibrated only up to halo masses of around $10^{12}\mathrm{M}_\odot$. It is possible that higher mass halos, in the rapid growth phase of their assembly, have a different distribution of orbital parameters for their merging halos. Indeed, as discussed by \cite{klypin_multidark_2016}, these highest mass halos form from the highest, most rare peaks of the density field, which tend to be more spherical than lower peaks. As such, the orbits of merging halos are expected to be more radial. \cite{klypin_multidark_2016} suggest this as an explanation for the upturn in the concentration--mass relation at the highest masses. This mass-dependence in the orbital parameter distribution of merging halos may therefore be a key missing ingredient in our model, and may at least partially explain the lower mean concentrations predicted by our model at the highest halo masses.

We do find that our model accurately predicts the correlation function of scale radii over time for these halos---indicating that it does capture the effects of the rapid growth phase of halo assembly on the evolution of scale radii.

Simulations of non-cold dark matter in which the power spectrum has a cut-off at low masses show a turnover in the concentration-mass relation. As has been shown by \cite{ludlow_mass-concentration-redshift_2016} such a turnover can be predicted by models which relate the concentration to the assembly history of halos. As such, we expect our model to capture this behavior. We intend to explore such scenarios and confirm this expectation in a future work.

While our model, which is implemented and available within the open source \href{https://github.com/galacticusorg/galacticus}{\textsc{Galacticus}} toolkit, offers further insight into the physics that determines halo concentrations, it also provides a practical method to assign concentrations to halos with merger histories derived from non-N-body means (e.g. those derived using Press-Schechter-based approaches). Furthermore, since our model is based on a simple physical principle, we expect it to be applicable beyond the cold dark matter scenario to which we have applied it here, in the same way that the \cite{ludlow_mass-concentration-redshift_2016} model performs well for warm dark matter scenarios also. This will be invaluable in computing the properties of halos and subhalos for a wide variety of non-CDM scenarios.

\section*{Acknowledgements}

Work by DG and TJ was supported by the Provost’s Office at Haverford College. TJ thanks the Carnegie Observatories for their support, and Gwen Rudie for organizing the summer intern program, which was funded in part by support from The Rose Hills Foundation and The Ralph M. Parsons Foundation, within which this work was carried out. AJB and DG thank Marc Kamionkowski for discussions and support during the development of the initial concept for this work, which was carried out with support from the Moore Foundation.

We thank Sownak Bose and Stelios Kazantzidis for valuable discussions. The calculations used in this paper were performed on the {\tt mies} cluster, made available through a generous grant from the Ahmanson Foundation. The MultiDark Database used in this paper and the web application providing online access to it were constructed as part of the activities of the German Astrophysical Virtual Observatory as result of a collaboration between the Leibniz-Institute for Astrophysics Potsdam (AIP) and the Spanish MultiDark Consolider Project CSD2009-00064. The Bolshoi and MultiDark simulations were run on the NASA’s Pleiades supercomputer at the NASA Ames Research Center. The MultiDark-Planck (MDPL) and the BigMD simulation suite have been performed in the Supermuc supercomputer at LRZ using time granted by PRACE.

\section*{Data availability}

The data underlying this article are available in Zenodo, at \href{https://doi.org/10.5281/zenodo.4277920}{\tt https://doi.org/10.5281/zenodo.4277920}, N-body simulation data from the COCO simulation is publicly available (after registration) at \href{https://cocos.ocean.icm.edu.pl/}{\tt https://cocos.ocean.icm.edu.pl/}. N-body simulation data from the VSMDPL simulation is publicly available (after registration) at \href{https://www.cosmosim.org/}{\tt https://www.cosmosim.org/}.

\bibliographystyle{aasjournal}
\bibliography{notes}

\appendix

\section{Unresolved Accretion}\label{app:unresolved}

Our merger trees have a finite mass resolution, below which accretion of halos is no longer resolved. Since those unresolved halos would contribute to the energy of the halo with which they merge, we must account for their contributions. We first develop an analytic estimate of the energy contributed by unresolved accretion, and then calibrate this model through a resolution study to ensure that our results are independent of merger tree resolution.

\subsection{Analytic Estimate}\label{app:unresolvedAnalytic}

We consider the energy provided by halos accreted below the mass resolution of our merger trees. On average unresolved halos of mass $M$ will have an orbital energy per unit mass of (see equation~\ref{eq:energyOrbital})
\begin{equation}
    \epsilon_\mathrm{orbit}(M) = \left[ -{\mathrm{G} M_1 \over R_1} + \frac{1}{2} {\langle v^2 \rangle \over 1+\mu} \right] (1+\mu)^{-\gamma},
    \label{eq:energyOrbitalUnresolved}
\end{equation}
where $\langle v^2 \rangle$ is the mean squared velocity of virial crossing orbits, and $\mu=M/M_1$ with $M$ being the mass of the unresolved halo, and $M_1$ is the mass of the most massive progenitor. 

In general we expect $\mu \ll 1$, but we can nevertheless account for this dependence. The mean energy per unit mass over all unresolved halos is found by averaging over the mass function:
\begin{equation}
 \langle \epsilon_\mathrm{orbit} \rangle = \left. \int_0^{M_\mathrm{res}} \epsilon_\mathrm{orbit}(M) M \frac{\mathrm{d}N}{\mathrm{d}M}(M) \mathrm{d}M \right/  \int_0^{M_\mathrm{res}} M \frac{\mathrm{d}N}{\mathrm{d}M}(M) \mathrm{d}M.
 \label{eq:energyOrbitalUnresolvedIntegral}
\end{equation}

Assuming a power-law halo mass function for the unresolved halos, $\mathrm{d}N/\mathrm{d}M \propto M^{-a}$ with $a\approx 1.9$ \citep[e.g.][]{springel_aquarius_2008}, we can use equation~(\ref{eq:energyOrbitalUnresolved}) in equation~(\ref{eq:energyOrbitalUnresolvedIntegral}) and evaluate the integral, from which we find:
\begin{equation}
    \epsilon_\mathrm{orbit} =  -C_\phi {\mathrm{G} M_1 \over R_1} + C_\mathrm{K}\frac{1}{2} {\langle v^2 \rangle \over 1+\mu_\mathrm{res}} ,
\end{equation}
where $\mu_\mathrm{res}=M_\mathrm{res}/M_0$, 
\begin{equation}
    C_\phi = (2-a) \mu_\mathrm{res}^{-2+a} B \left( {\mu_\mathrm{res} \over 1+\mu_\mathrm{res}}; 2-a, -2+a+\gamma \right),
\end{equation}
and
\begin{equation}
    C_\mathrm{K} = (2-a) \mu_\mathrm{res}^{-2+a} B \left( {\mu_\mathrm{res} \over 1+\mu_\mathrm{res}}; 2-a, -1+a+\gamma \right),
 \label{eq:energyUnresolvedOrbital}
\end{equation}
where $B(x;a,b)=\int_0^x t^{a-1} (1-t)^{b-1} \mathrm{d}t$ is the incomplete beta-function.

For the internal energy of each unresolved halo we can write:
\begin{equation}
    E_\mathrm{int}(M) = - s(M) {{\mathrm G} M^2 \over R} (1+\mu)^{-\gamma}
\end{equation}
where $s(M)$ depends weakly on mass via the mass dependence of the density profile. If we assume that all accreted halos have the same virial density, then $R = (3 M / 4 \pi \bar \rho)^{1/3}$, so that
\begin{equation}
    E_\mathrm{int}(M) = - s(M) {\mathrm G} \left( {4 \pi \bar\rho \over 3} \right)^{1/3}M^{5/3} (1+\mu)^{-\gamma},
\end{equation}
then the mean internal energy per unit mass of accreted subresolution halos is
\begin{equation}
\epsilon_\mathrm{int} = \left. \int_0^{M_\mathrm{res}} E_\mathrm{int}(M) M^{-a} \mathrm{d}M \right/  \int_0^{M_\mathrm{res}} M M^{-a} \mathrm{d}M.
\end{equation}

From an analysis of halos generated in higher resolution merger trees we find that $s(M) \propto M^{-c}$ with $c \approx 0.03$. The internal energy per unit mass of accreted subresolution halo is then
\begin{eqnarray}
 \epsilon_\mathrm{int} &=& (2-a-c) / (8/3 - a-c) \nonumber \\
 & & \times _2F_1\left[(\gamma,8/3-a-c),(11/3-a-c),-\mu_\mathrm{res}^{-1}\right] \nonumber \\
 & & \times E_\mathrm{int}(M_\mathrm{res}) / M_\mathrm{res},
 \label{eq:energyUnresolvedInternal}
\end{eqnarray}
where $_{2}F_{1}(a,b;c;z)=\sum _{n=0}^{\infty }{\frac {(a)_{n}(b)_{n}}{(c)_{n}}}{\frac {z^{n}}{n!}}.$ is a hypergeometric function \citep[][\S15.1.1]{abramowitz_handbook_1970}.

The total energy contributed by unresolved accretion is then taken to be
\begin{equation}
    E_\mathrm{unres} = \beta \left[\epsilon_\mathrm{int}+\epsilon_\mathrm{orb}\right] (1+b) M_\mathrm{unres},
\end{equation}
where $M_\mathrm{unres}$ is the increase in mass of a halo which \emph{is not} accounted for by resolved progenitor halos, $b$ is a parameter of our model (introduced in \S\ref{sec:model}) which represents an overall boost in the energy of a halo relative to its progenitors, and $\beta\sim 1$ is a parameter which we introduce to allow us to calibrate this analytic estimate of the energy of unresolved halos.

The parameter $\beta$ introduced above should be chosen such that the energies (and, therefore, concentrations) of halos are insensitive to changes in the resolution of the merger tree used to characterize the halo's assembly. In the next subsection we will determine an appropriate value for $\beta$ through a resolution study.

\subsection{Resolution Study}\label{app:unresolvedResStudy}

\begin{figure}
\begin{center}
 \includegraphics[width=85mm]{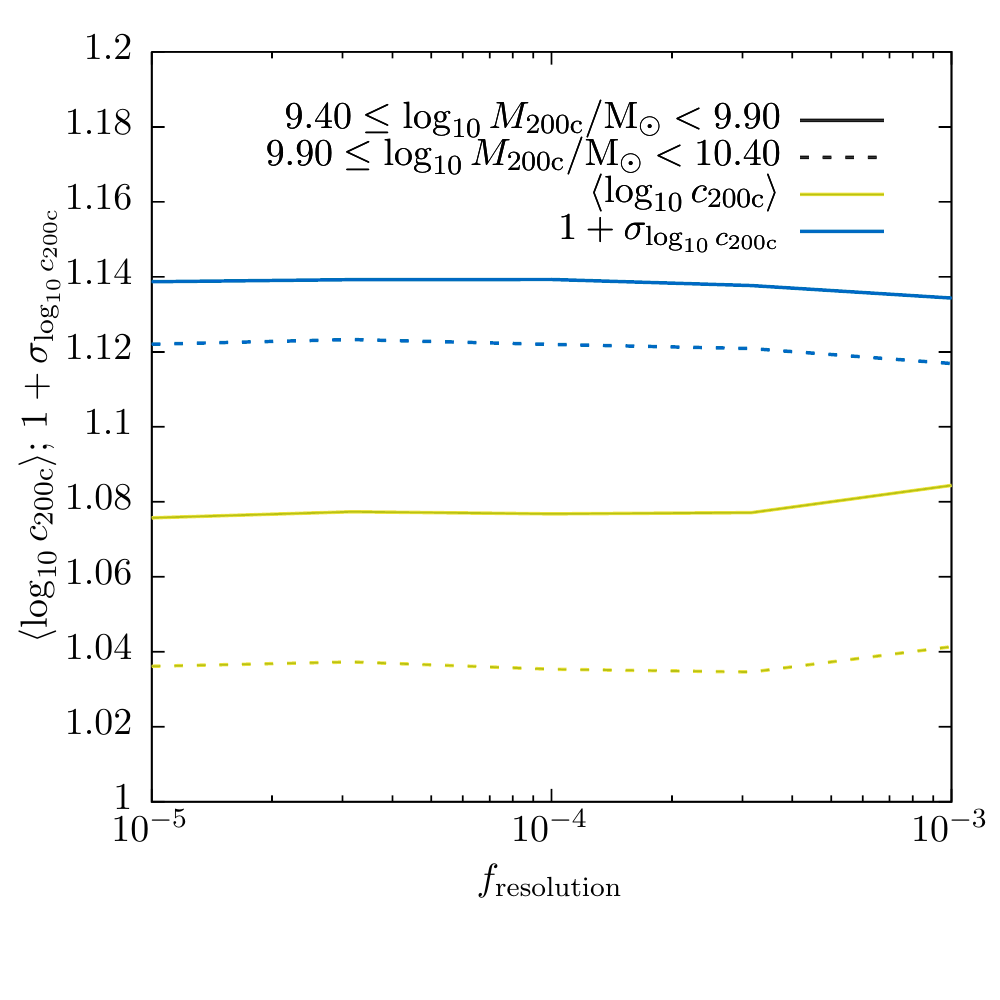}
 \end{center}
 \caption{The mean (yellowish-green line) and root-variance (blue line) of the distribution of concentrations for $z=0$ halos in two mass ranges (as indicated by line type) as a function of the merger tree resolution. Results are shown for $\beta=0.55$ which we find gives results that are stable with merger tree resolution.}
 \label{fig:resolutionConvergence}
\end{figure}

In the above we introduced a parameter, $\beta$, which multiplies our analytic estimate of the energy contributed by unresolved halos. The value of $\beta$ should be of order unity, and should be chosen such that the mean concentration (at a given halo mass) predicted by our model is independent of the chosen merger tree mass resolution. By exploring different values of $\beta$ and applying our model to merger trees constructed with different mass resolutions we find that a value of $0.55$ gives stable results, as shown in Fig.~\ref{fig:resolutionConvergence}.

The yellowish-green line in Fig.~\ref{fig:resolutionConvergence} shows the mean concentration predicted (for halos in the mass range indicated above each panel) as a function of the mass resolution parameter, $f_\mathrm{resolution}$. The results are essentially independent of resolution, indicating that our model for unresolved accretion works.

It should be noted that the analytic estimate of the energy contributed by unresolved accretion in \S\ref{app:unresolvedAnalytic} estimates only the mean of this contribution---we do not attempt to estimate the scatter in this quantity. However, the blue lines in Fig.~\ref{fig:resolutionConvergence} show that the scatter in concentration is also independent of merger tree mass resolution, even though we neglect the scatter contributed by unresolved accretion. Since the mass function of accreted halos rises steeply with decreasing halo mass, the law of large numbers implies that the actual energy contributed by unresolved accretion onto a halo will be close to the mean expectation.

\section{Posterior distribution}\label{sec:posterior}

\begin{figure}
 \newcommand{\triangledir}{.}
\renewcommand{\arraystretch}{0}
\setlength{\tabcolsep}{0pt}
\begin{tabular}{l@{}c@{}r@{}l@{}c@{}r@{}}
\multicolumn{3}{c}{\includegraphics[scale=1.0]{\triangledir/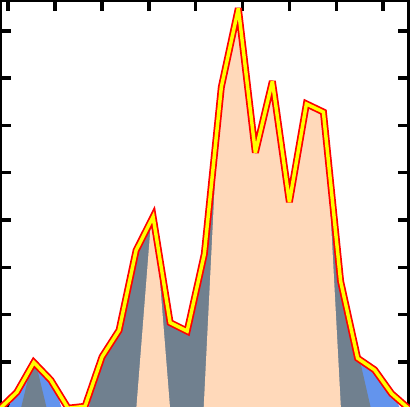}}&\multicolumn{3}{c}{\includegraphics[scale=1.0]{\triangledir/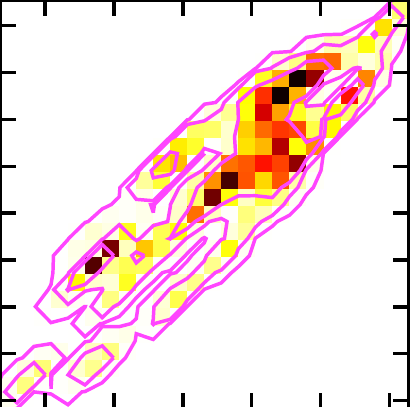}}\\
\multicolumn{1}{p{37.5384615384615pt}}{\raisebox{121pt-\widthof{\normalsize x}-\widthof{\normalsize $1.4$}}[0pt][0pt]{\rotatebox{90}{\normalsize $1.4$}}}&\multicolumn{1}{p{37.5384615384615pt}}{\hspace{19.6774193548387pt}\raisebox{121pt-\widthof{\normalsize x}-\widthof{\normalsize $ \gamma$}}[0pt][0pt]{\rotatebox{90}{\normalsize $ \gamma$}}}&\multicolumn{1}{r}{\raisebox{121pt-\widthof{\normalsize x}-\widthof{\normalsize $1.6$}}[0pt][0pt]{\rotatebox{90}{\normalsize $1.6$}\hspace{3pt}}}&\multicolumn{3}{c}{\includegraphics[scale=1.0]{\triangledir/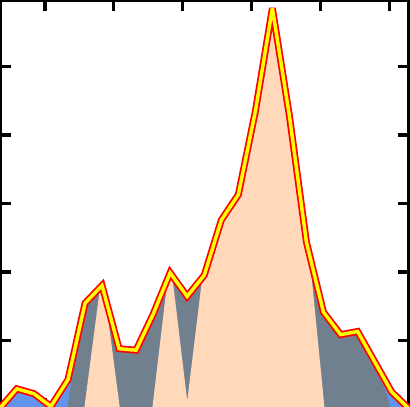}}\\
&&&\multicolumn{1}{p{37.5384615384615pt}}{\raisebox{0pt-\widthof{\normalsize $6.3 \times 10^{-1}$}}[0pt][0pt]{\rotatebox{90}{\normalsize $6.3 \times 10^{-1}$}}}&\multicolumn{1}{p{37.5384615384615pt}}{\hspace{19.6774193548387pt}\raisebox{0pt-\widthof{\normalsize $ b$}}[0pt][0pt]{\rotatebox{90}{\normalsize $ b$}}}&\multicolumn{1}{r}{\raisebox{0pt-\widthof{\normalsize $6.9 \times 10^{-1}$}}[0pt][0pt]{\rotatebox{90}{\normalsize $6.9 \times 10^{-1}$}\hspace{3pt}}}\\
\end{tabular}

 \vspace{15mm}
 \caption{The posterior distribution over the model parameters $b$ and $\gamma$. (Nuisance parameters are not shown.) The off-diagonal panel shows the posterior distribution over both model parameters, while on-diagonal panels show the posterior distribution over individual model parameters. In the off-diagonal panel, colours show the probability density running from white (low probability density) to dark red (high probability density). Contours are drawn to enclose 99.7\%, 95.4\%, and 68.3\% of the posterior probability when ranked by probability density (i.e. the highest posterior density intervals). In on-diagonal panels the curve indicates the probability density. Shaded regions indicate the 68.3\%, 95.4\%, and 99.7\% highest posterior density intervals.}
 \label{fig:modelPosterior}
\end{figure}

Figure~\ref{fig:modelPosterior} shows the posterior distribution over the model parameters determined from our MCMC simulation described in \S\ref{sec:model}, with nuisance parameters not shown. Both parameters are well-constrained by the N-body data. It is also apparent that their values are strongly correlated in the posterior distribution.

\section{Effects of Choice of Density Profile}\label{sec:profile}

Throughout this work we have made use of the NFW \citep{navarro_universal_1997} density profile for dark matter halos. More recent works \citep{navarro_inner_2004,gao_redshift_2008,navarro_diversity_2010} have suggested that the \cite{einasto_construction_1965} profile is a more accurate description of the density profiles of cosmological cold dark matter halos. While the NFW profile is described by just a single parameter, $r_\mathrm{s}$, (once the total mass and density contrast of the halo are fixed), the Einasto profile requires two parameters, the radius at which the logarithmic slope of the density profile equals $-2$, $r_{-2}$, and a shape parameter, $\alpha$.

Our model allows us to uniquely determine the density profile of a halo only in the case of a single-parameter family, such as NFW. Therefore, to explore a two-parameter family such as the Einasto profile we fix the value of the shape parameter $\alpha$ using the fitting function of \cite{gao_redshift_2008}. We then apply our model to the case of Einasto profiles. We use the most probable \emph{a posteriori} values of the parameters $b$ and $\gamma$ (and all nuisance parameters) determined from our MCMC simulation. Since that simulation utilized NFW halos we may expect some offset in the results when applied to Einasto profiles.

\begin{figure*}
\begin{tabular}{cc}
 \includegraphics[width=85mm]{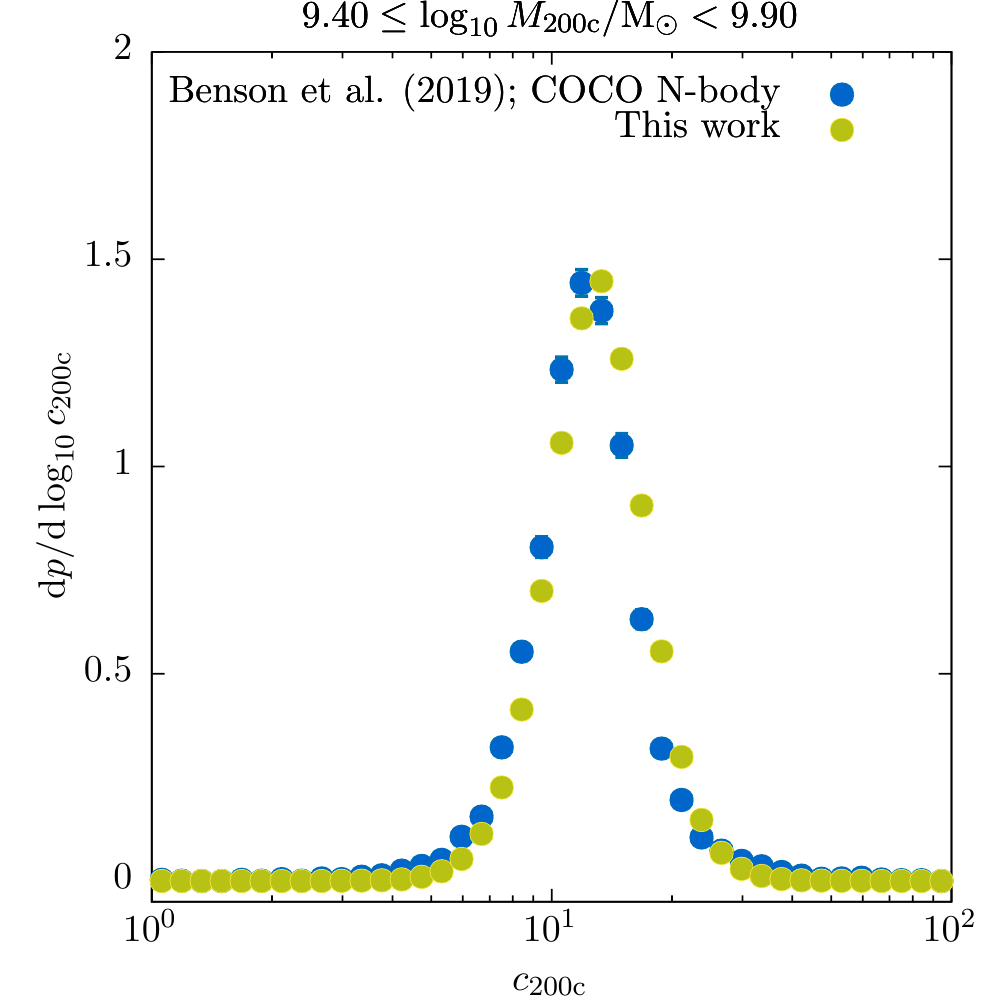} &  \includegraphics[width=85mm]{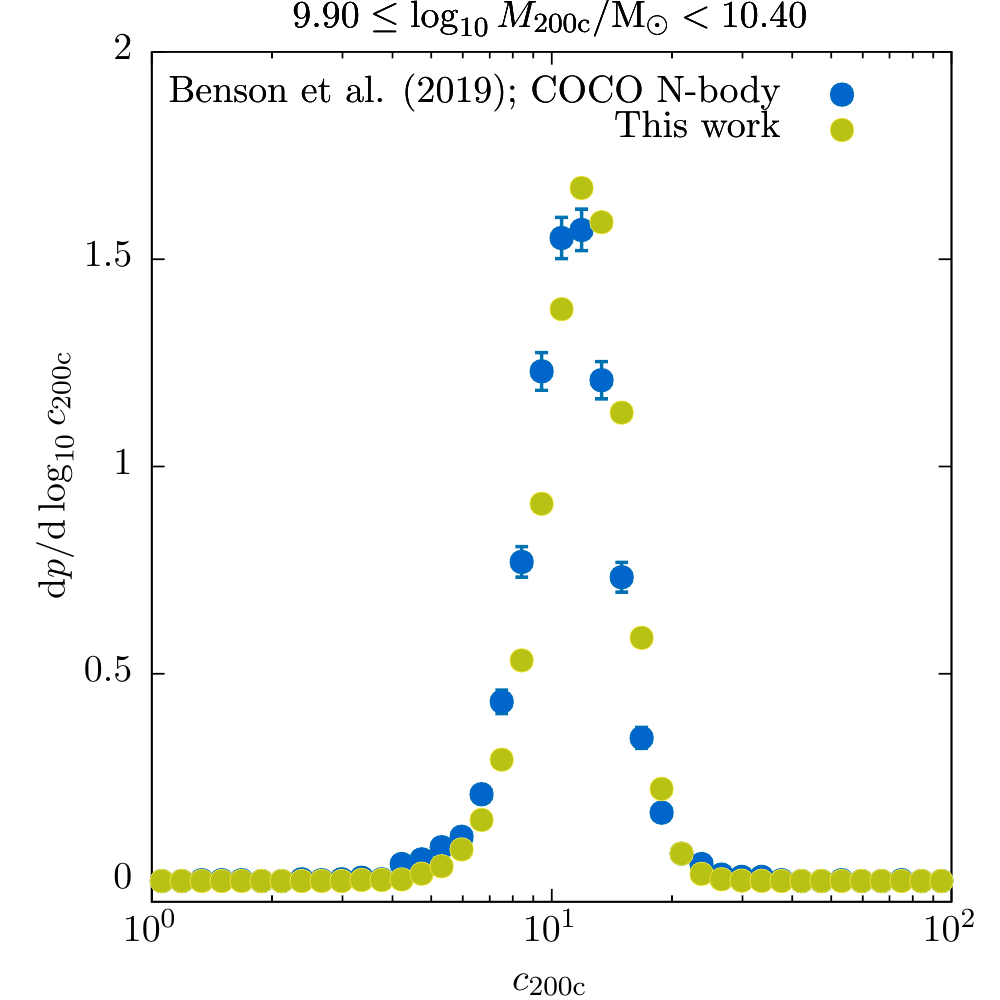}
 \end{tabular}
 \caption{Distributions of concentration parameter for $z=0$ halos in two mass ranges (as indicated above each panel) using an Einasto density profile in our model. Blue points show the measured distribution from the COCO N-body simulation from \protect\cite{benson_halo_2019}, while yellowish-green points show results from the maximum likelihood model in this work.}
 \label{fig:distributionEinasto}
\end{figure*}

\begin{figure}
\includegraphics[width=85mm]{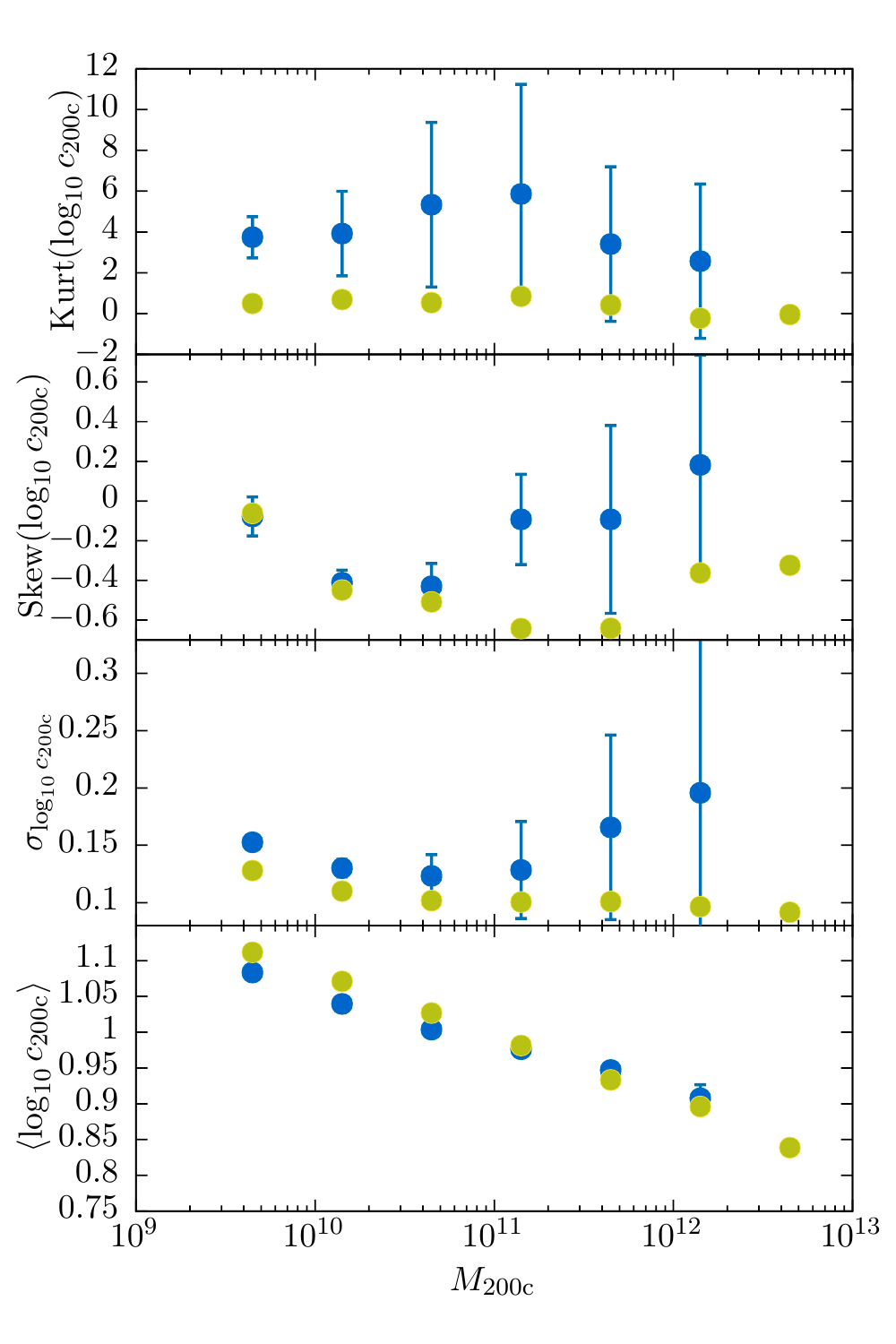}
\caption{Moments of the distribution of concentrations, under the assumption of Einasto halo profiles. Each panel shows a different moment (mean, scatter (i.e. root-variance), skewness, and (excess) kurtosis from bottom to top) as a function of $z=0$ halo mass, with results from this work shown as yellowish-green points, and from the COCO N-body measurements from \protect\cite{benson_halo_2019} as blue points. For the N-body results uncertainties arising from the finite number of halos available are found via bootstrapping from the original measurements. Uncertainties due to the finite number of halos used in this work are much smaller and so are not shown.}
\label{fig:statsEinasto}
\end{figure}


Fig.~\ref{fig:distributionEinasto} shows distributions of concentration parameters for this calculation. There is a small but clear shift to higher concentrations compared to the NFW profile case, while the scatter in concentration is slightly reduced, as can be seen more clearly in Figure~\ref{fig:statsEinasto} which shows the moments as of the concentration as a function of halo mass under the assumption of Einasto profiles. Our qualitative conclusions are therefore not affected by the choice of density profile. The quantitative agreement with N-body results when using an Einasto profile could be improved by performing a new MCMC simulation to recalibrate $b$ and $\gamma$.

\end{document}